%% file: memo_signals_ramo.tex
\documentclass[10pt]{article}
\usepackage[twoside,pdftex,papersize={210mm,280mm}, total={16.8cm,23.6cm}, left=1.5cm, top=1.5cm, right=1.5cm, bottom=2.5cm, includehead=true]{geometry}
\usepackage{extarrows}

\usepackage{amstext}
\usepackage{amsmath}
\usepackage{amssymb}
\usepackage{amsfonts}
\usepackage{graphicx}  
\usepackage{enumerate}
\usepackage{float} 
\usepackage{textcomp}
\usepackage{upgreek}
\usepackage{rotating}
\usepackage{multirow}
\usepackage{xspace}
\usepackage{color}
\usepackage{mathtools}
\usepackage{braket}
\usepackage{fancybox,calc}
\usepackage{bm}
\usepackage{cancel}
\usepackage{sty/authblk}
\usepackage{ctable}
\usepackage{placeins}
\usepackage{mciteplus}
\usepackage[hidelinks]{hyperref}
\usepackage{appendix}
\usepackage{color}
\usepackage{colortbl}
\newboolean{pdflatex}
\setboolean{pdflatex}{false} 
\mathtoolsset{showonlyrefs=true}
\mathtoolsset{showmanualtags=true}

\newcommand{\etal}{{\it et al.}}
\newcommand{\fr}{\dfrac}
\newcommand{\dr}{\drac}
\newcommand{\dsps}{\displaystyle}
\newcommand{\drac}[2]{\dsps \frac{\mathrm{d} #1}{\mathrm{d} #2}}
\newcommand{\prac}[2]{\dsps \frac{\partial #1}{\partial #2}}
\newcommand{\pr}{\prac}

\newcommand{\executeiffilenewer}[3]{%
\ifnum\pdfstrcmp{\pdffilemoddate{#1}}%
{\pdffilemoddate{#2}}>0{\immediate\write18{#3}}
\fi}

\IfFileExists{InkscapeCommand.tex}%
{}
{}

\graphicspath{{figures/}}

\definecolor{lightgray}{gray}{0.95}

\begin{document}

\title{Signal Formation in Various Detectors}
\renewcommand\Authands{, } 
\renewcommand\Affilfont{\itshape\small} 

%

\author[]{Manolis Dris}
\author[]{Theo Alexopoulos}

\affil[]{National Technical University of Athens, Department of Physics,\\
9 Heroon Polytechniou Street, GR 157 80, Athens, Greece}


\maketitle

\begin{abstract}
\noindent
In this write-up we present the general theory of the signal formation in various detectors. We follow a pedagogical analysis 
and presentation such that the results could be easily understood 
and applied by the interested reader to the different 
detector configurations.
We include few applications to gaseous detectors, namely,
 Monitored Drift Tubes (MDT) and micro-pattern gaseous detectors of the  Micromegas type.
\end{abstract}

\tableofcontents


\input{general_theory}

\input{Acknowledgments}

\input{bibliography}

\end{document}

%% file: general_theory.tex
\section{Signal formation in a small-size detector} \label{sec:therory}

Let us consider a detector with small dimensions such that wave propagation effects are not important. 
This means that for the larger dimension, $d_\mathrm{max}$, of the detector we have $d_\mathrm{max} \ll \lambda_\mathrm{min}=c'/f_\mathrm{max}$, where $f_\mathrm{max}$ is the maximum frequency needed to describe the detector signal, $c'$ is the electromagnetic wave velocity in the detector.

\noindent The situation is semi-static. 
This means that even though there is time dependence, the time rates of change at any time are so small that the relations of electrostatics hold.

\noindent We will prove a generalization of the Shockley-Ramo theorem ~\cite{ramo, shockley}. 
This generalization is similar to the one described in reference ~\cite{gatti}. 
The Shockley-Ramo theorem refers to signals induced on grounded conductors due to moving charges in the space between the conductors. 
In the generalization, the conductors need not be grounded; they may have different potentials and they are connected to an external circuit (network).

\begin{figure}[htb]
  \centering
  \input{#figures/outerlattice.pdf_tex}
  \caption{System of $N+1$ conductors and an external network to which they are connected. 
  There is a time-varying charge distribution, $\rho(\bm{x},t)$, in the space between the conductors. The charge velocity is $u(\bm{x},t)$.}\label{fig:mmsr-1}
\end{figure}
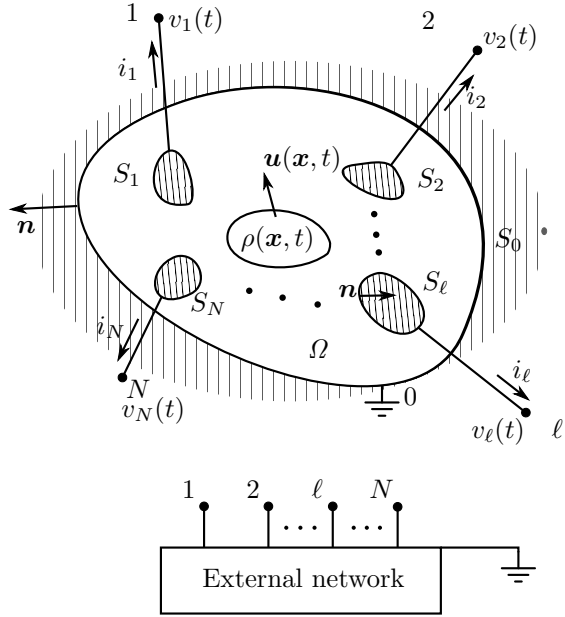

\noindent Let us consider the situation shown in Figure \ref{fig:mmsr-1}, where there are $N$, internal, ideal conductors.
The conductors are of finite extend. 
In addition, there is an external conductor $0$ surrounding all the internal conductors. 
This external conductor is held at a known reference potential, which we assume to be zero. 
The external conductor can be considered to be extended, totally or partially, to infinity.

\noindent We examine the case when the space between the conductors contains a linear dielectric material with permittivity (dielectric function) $\varepsilon (\bm{x}) =\varepsilon_\mathrm{r} (\bm{x}) \varepsilon_\mathrm{o}$, where $\varepsilon_\mathrm{r}$ is the relative permittivity and $\varepsilon_\mathrm{o}$ is the vacuum permittivity or electric constant. 
Thus, the dielectric material might not be homogeneous, on the other hand the dielectric function does not depend on frequency.

\noindent The shaded part of Figure \ref{fig:mmsr-1} denotes the space inside the conductors. 
The space between the conductors is considered as the external space of the conductors and it is denoted by $\it\Omega$. 
This space is multiply connected, so the total surface $S$ surrounding this volume is the union of all the surfaces of the individual conductors
\[
S=S_0 \cup S_1\cup S_2 \ldots \cup S_N = \cup_{i=0}^N S_i
\]
The charge distribution $\rho$ is, in general, a function of space $\bm{x}$ and time $t$, so $\rho = \rho(\bm{x},t)$. 
The velocity of the charge distribution is also a function of space and time, $\bm{u} = \bm{u} (\bm{x},t)$. 
In this analysis, these two functions, $\rho(\bm{x},t),\ \bm{u} (\bm{x},t),$ are supposed to be known, given, functions.

\noindent The signals are taken from the conductors $1,2,\ldots,N$ of the detector and are then fed to the external network. 
Usually, the detector is characterized as an $N$ conductor (signal) detector, ignoring the reference conductor $0$. 
 $\bm{u} (\bm{x},t)$ is the drift velocity of the moving charges that are created inside the detector. 
The drift velocity, in general, is a function of the electric field, $\bm{E}=\bm{E}(\bm{x},\ t)$.  
This electric field depends on the external potentials, the bias potentials of the detector, and on the space charges created inside the detector. 
In addition, if there is an external magnetic field $\bm{B}$, the drift velocity will be affected. 
Usually, the bias voltages are constant in time and furthermore the (external) field they produce is much stronger than the (internal) field produced by the space charges.
In this case, the space charges create a negligible ``internal'' electric field. 
In some cases, we might have a large concentration of charges; as a result, a strong ``internal'' electric field is created, and must be taken into account, appropriately.  
To simplify our calculations, we will consider that the biasing potentials and space charges produce fields independent of time, thus we can denote the velocity as a function of space only, $\bm{u} = \bm{u} (\bm{x})$.

\noindent Inside the detectors, along with the creation of electrons, positive charges are also created. 
In the case of gaseous detectors, this is due to the ionization process. 
At each ionization point, the negative charges (primarily electrons) and the positive charges (ions) move in opposite directions. 
The absolute value of the drift velocity of the electrons is much larger than the drift velocity of the ions, $u_\mathrm{e} \gg u_\mathrm{ion}$ (for Ar gas at normal conditions, $u_\mathrm{e}\approx 200u_\mathrm{ion}$). 
Evidently, the corresponding currents of the positive and negative charges have the same direction.  
In order to find the total result, we need to apply the superposition principle, taking into account all the contributing charge types. 
We will examine the case where we consider only one type of charges.  
We should mention here that one could argue that in this approach, conservation of charge is not taken into account.  
In reality, when an amount of negative charge is created, an equal amount of positive charge is produced at the same time.
One could imagine that an equal amount of charge is produced and remains static at the point of ionization. 
Another argument is that because we treat the problem as a semi-electrostatic problem, there is no use of the full Maxwell equations which contain the conservation of charge.

\noindent Let us assume that there is an electrostatic condition at any given moment (Figure \ref{fig:mmsr-1}). 
We apply the divergence theorem ~\cite{gatti, jackson, blum}, which leads to Green's reciprocity theorem.
This can be proven for a linear dielectric, even when it is not homogeneous, in which case permittivity $\varepsilon = \varepsilon (\bm{x}) $ depends on the position $\bm{x}$.

\noindent The reciprocity theorem can be expressed as follows:
\begin{equation}
  \int_{\it\Omega} \it\Phi ' \rho \mathrm{d}^3x + \int_S \it\Phi ' \sigma \mathrm{d}a = \int_{\it\Omega} \it\Phi  \rho ' \mathrm{d}^3 x + \int_S \it\Phi \sigma ' \mathrm{d} a
  \label{eq:mmsr-1-1}
\end{equation}
Eq. \eqref{eq:mmsr-1-1} states that the volume distribution of free charge, $\rho $, in three dimensional space $\it\Omega$ and the surface charge distribution, $\sigma $, on the surface $S$ surrounding the above space, create a potential $\it\Phi $ inside $\it\Omega$.
Instead of a continuous distribution, one can use a free point charge. 
This is a mathematically simpler case to treat. 
Starting from the point charge analysis, one could superimpose the appropriate partial results of many point charges to get the result for the continuous charge distribution. 
We can use the continuous charge distribution result to obtain the moving discrete point charge result, by introducing the appropriate delta function distribution.

\noindent We will proceed with the generalization of the Shockley-Ramo theorem, for a continuous charge distribution which changes with time.

\noindent The charge distribution shown in Figure \ref{fig:mmsr-1}, in general, dependents on time. 
The rate of change is sufficiently small. 
There is an external network connected to the conductors of the detector and to the surrounding reference conductor, $0$. 
The actual (instantaneous) semi-electrostatic state of Figure \ref{fig:mmsr-1} is described at time $t$ by the charge and potential quantities by
\begin{equation}
  \left[ \rho (\bm{x}, t), q_0(t), q_1(t), q_2(t), \ldots, q_N(t) \right], \quad \left[ {\it\Phi} ({\bm{x}, t)}, 0, v_1(t), v_2(t), \ldots, v_N(t) \right]
  \label{eq:mmsr-1-3}
\end{equation}
The running indices refer to quantities related to the corresponding conductors, and $\bm{x}$ is the position vector for points inside $\it\Omega $, the space between the conductors.

\noindent Let us now imagine $N$ electrostatic states of the $N$ conductors of the detector. 
We use two indices: index $j=1, 2, \ldots, N$ describes the state, while index $l=1, 2, \ldots, N$ describes the conductor. 
Here, the potentials and fields are constant and there is no charge inside the space $\it\Omega $, i.e. $\rho (\bm{x}, t)=0$. 
These states are shown by
\begin{equation}
  \left[ 0, Q_{j1}, \ldots, Q_{jl}, \ldots,Q_{jN} \right], \quad \left[ {\it\Phi _j} (\bm{x}), 0,0,\ldots, V_{l(=j)}, \ldots ,0\right]\qquad j=1, 2, \ldots, N
  \label{eq:mmsr-1-4}
\end{equation}
In Eq. \eqref{eq:mmsr-1-4}, $Q_{jl}$ is the charge of conductor $l$ in state $j$, where only conductor $l=j$ has a non-zero potential, $V_{l(=j)} \neq 0$, while all the others have a zero potential, $V_{l(\neq j)}=0$. ${\it\Phi _j}(\bm{x})$ is the potential in space $\it\Omega $ corresponding to state $j$.

\noindent We apply Eq. \eqref{eq:mmsr-1-1} for the actual state described by Eqs. \eqref{eq:mmsr-1-3}, and state $j$, described by Eq. \eqref{eq:mmsr-1-4}, and we get
\begin{equation}
  \int_{\it\Omega}  {\it\Phi _j}(\bm{x})\rho (\bm{x}, t) \mathrm{d}^3 x+ V_j \int_{S_j} \sigma _j (\bm{x}, t) \mathrm{d}a = \sum_{l=1}^{N} v_l(t) Q_{jl}, \quad j=1, \ldots, N
  \label{eq:mmsr-1-5}
\end{equation}
Evidently, the charge of conductor $j$ is $q_j(t)= \int_{s_j} \sigma _j (\bm{x}, t)  \mathrm{d}a$.
We define $c_{jl}$ as
\begin{equation}
  c_{jl} = \frac{Q_{jl}}{V_j}
  \label{eq:mmsr-1-6}
\end{equation}
therefore, the following equations stand at any given moment
\begin{equation}
  \phi _j(t)= \frac{1}{V_j} \int_V {\it\Phi _j}(\bm{x}) \rho  (\bm{x}, t) \mathrm{d}^3 x = -q_j(t) + \sum_{l=1}^{N} c_{jl} v_l (t), \quad j=1, \ldots, N
  \label{eq:mmsr-1-7}
\end{equation}
The physical meaning of the various quantities is understood from the definitions of states of Eqs. \eqref{eq:mmsr-1-3}, \eqref{eq:mmsr-1-4} and Eq.~\eqref{eq:mmsr-1-6}. 
Specifically, ${\it\Phi _j}(\bm{x})$ is the potential in the space between the conductors, when there is no free charge while only conductor $j$ has a constant potential $V_j$, and all the other conductors have zero potential. 
$c_{jl}$ is the quotient of the free charge induced on conductor $l$, (when conductor $j$ has potential $V_j$, while all the rest have zero potential, and there is no charge in the space between them) divided by $V_j$. 
For the coefficients  we have $c_{ii}>0$ and $c_{ij}\leq 0, \ i\neq j$. 
These quantities are often referred to with different names, one name is \textsl{electrostatic inductance coefficients}. 
They have the dimension of a capacity but they don't have an electrotechnical representation; as a matter of fact, some of them are negative.

\noindent We will now see that we can introduce capacitances that have the usual physical meaning as the capacitances in electrotechnology. 
This allows us to use the usual methods of electrical circuit theory for solving the problems of detector signals.

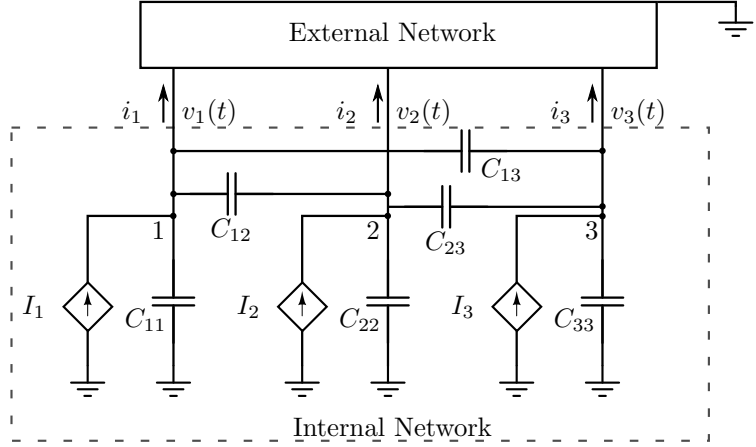
\begin{figure}[htb]
 \centering
 \input{#figures/equivcirc.pdf_tex}
 \caption{The equivalent detector circuit with three conductors and an external circuit.}\label{fig:mmsr-1-2}
\end{figure}

\noindent Evidently, from Eq. \eqref{eq:mmsr-1-7}, when $\rho (\bm{x}, t)=0$, we arrive at Eq. \eqref{eq:mmsr-1-8} for the charges and the potentials of the conductors. 
Since we have electrostatic states, we replaced \(q\)'s by $Q$ and \(v\)'s by $V$
\begin{gather}\label{eq:mmsr-1-8}
  \begin{aligned}
    Q_j &= \sum_{l=1}^{N} c_{jl} V_l, \quad j=1, \ldots, N \\
    U &= \dfrac{1}{2} \sum_{i,j=1}^{N} c_{ij} V_iV_j
  \end{aligned}
\end{gather}
where $U$ is the electrostatic energy of the system, $Q_j$ is the charge induced in conductor $j$ by all the conductors when they have potentials $V_l$, $l=1, 2, \ldots, N$. 
In order to introduce the ``usual'' capacitances, we need to define the potential differences between the various nodes. 
For this purpose, we successively add to and subtract from the second members of Eqs. \eqref{eq:mmsr-1-8} the sums $(\sum_{l=1}^{N} c_{jl})V_j$. 
This yields 
\begin{equation}
  Q_j = C_{jj} V_j + \sum_{l=1}^{N}  C_{jl} (V_j-V_l), \quad j=1, \ldots, N
  \label{eq:mmsr-1-9}
\end{equation}
where
\begin{gather}\label{eq:mmsr-1-10}
  \begin{aligned}
    c_{jl} &= C_{lj}= - c_{lj} = -c_{jl} \quad \forall j\neq l, \ C_{ll}= \sum_{j=1}^{N} c_{lj}   \\
    c_{ij} &= C_{ij} \quad \forall i\neq j, \quad c_{ii}= \sum_{k=1}^{N} C_{ik}
  \end{aligned}
\end{gather}
The coefficients $C_{jl}$ are usually referred to as \textsl{capacity coefficients} and are all non-negative, $C_{jl}\geq 0$. 
Coefficients $C_{jj}$ represent the capacity between point (node) $j$ and reference conductor 0. 
Coefficients $C_{ij}$ are the usual capacities describing electrostatic coupling (interaction) between the corresponding conductors of the detector. 
These capacities are also called \textsl{two terminal capacitances}.
In this case, it is noteworthy that, according to Eq. \eqref{eq:mmsr-1-9}, for each coefficient $C_{jl}$ we have
\begin{equation}
  C_{jl} = - \frac{Q_j}{V_l}, \quad j\neq l
  \label{eq:mmsr-1-11}
\end{equation}
Remember that $Q_j$ is the induced charge on conductor $j$, when all other conductors have zero potential, $V_k=0 \ \forall k\neq l$, and only conductor $l$ has a non-zero potential, $V_l \neq 0$. 
For the remaining coefficients we have
\begin{equation}
  C_{jj} = \frac{Q_j}{V_j}
  \label{eq:mmsr-1-12}
\end{equation}
Notice that $Q_j$ is the charge induced in conductor $j$ when all other conductors have the same potential as $j$, namely $V_l = V_j$, for every  $l = 1, 2, \ldots, N$. 
Again, we follow the procedure above, i.e. we add to and subtract from the second member of Eq. \eqref{eq:mmsr-1-7} the sum $(\sum_{l=1}^N c_{jl}) v_j$ and, by combining this with the last term ($\sum_{l=1}^{N}$), we get
\begin{equation}
  \phi _j(t)= \frac{1}{V_j} \int_V {\it\Phi _j}(\bm{x}) \rho (\bm{x}, t) \mathrm{d}^3x = -q_j(t) + C_{jj}v_j(t)+\sum_{l=1}^{N}  C_{jl} \left( v_j(t)- v_l(t) \right)\quad j=1, \ldots, N
  \label{eq:mmsr-1-13}
\end{equation}
This leads us to the equivalent electrotechnical circuit of the detector, which is shown in Figure \ref{fig:mmsr-1-2} together with the external circuit and current sources. 

\noindent To determine the instantaneous currents and potentials, we will apply Eq. \eqref{eq:mmsr-1-7} or Eq. \eqref{eq:mmsr-1-13} at two nearby times. 
The charges and the potentials in general change with time, so if we take the time derivatives of these equations, we get
\begin{gather}\label{eq:mmsr-1-14}
  \begin{aligned}
    \drac{\phi_j (t)}{t} = \dr{}{t} \left\{ \frac{1}{V_j} \int_{\it\Omega}  {\it\Phi _j} (\bm{x}) \rho (\bm{x}, t) \mathrm{d}^3 x \right\}&= - \dr{q_j(t)}{t}+ \sum_{l=1}^{N}  c_{jl} \dr{v_l(t)}{t} \\
    &= - \dr{q_j(t)}{t}+ C_{jj} \dr{v_j(t)}{t} + \sum_{l=1}^{v_j(t)} C_{jl} \dr{\left(v_j(t)- v_l(t)\right)}{dt}
  \end{aligned}
\end{gather}
To find the time derivative of the integral, we apply the transport theorem (it is used in fluid mechanics), so we find
\begin{align}
  \dr{}{t} \left\{ \fr{1}{V_j} \int_{\it\Omega}  {\it\Phi _j} (\bm{x}) \rho (\bm{x}, t) \mathrm{d}^3 x \right\} &= \\
  \fr{1}{V_j} \int_{\it\Omega} \pr{ \left\{ {\it\Phi _j}(\bm{x}) \rho (\bm{x}, t) \right\}}{t} \mathrm{d}^3 x &+\fr{1}{V_j} \int_{S_j} {\it\Phi _j}(\bm{x}) \rho (\bm{x}, t) \left[ \bm{u} (\bm{x})\cdot \bm{n}(\bm{x} ) \right]\mathrm{d}a, \ \ j=1, \ldots, N
  \label{eq:mmsr-1-15}
\end{align}
It often happens that part of the charge distribution is constant with time (semiconductor detectors is one such case). 
In this case, one can note that only the part of the charge distribution which is moving contributes to the signal formation. 
We could say that the charge contributes to the small or large modification of the biasing static electric field, while the current density is only due to the moving charges. 
Using Eq. \eqref{eq:mmsr-1-15}, we get
\begin{align}
  \bm{J} (\bm{x}, t) &= \rho  (\bm{x}, t) \bm{u}(\bm{x}) \\
  \bm{\nabla} \cdot \bm{J} (\bm{x}, t) + \prac{\rho  (\bm{x}, t)}{t}&= 0 \\
  \drac{\phi_j(t)}{t}&= \fr{1}{V_j}\int_{\it\Omega} {\it\Phi _j}(\bm{x}) \prac{\rho (\bm{x}, t) }{t}\mathrm{d}^3x+\fr{1}{V_j} \int_{S_j}{\it\Phi _j}(\bm{x})\rho (\bm{x}, t) \left[ \bm{u}(\bm{x})\cdot \bm{n}(\bm{x}) \right]\mathrm{d}a \\
  &= -\int_{\it\Omega}  {\it\Phi _j}(\bm{x})\bm{\nabla}\cdot \bm{J} (\bm{x}, t) \mathrm{d}^3x+\fr{1}{V_j} \int_{S_j} {\it\Phi _j}(\bm{x}) \rho (\bm{x}, t) \left[ \bm{u} (\bm{x})\cdot \bm{n}(\bm{x}) \right]\mathrm{d}a
  \label{eq:mmsr-1-16}
\end{align}
Using the vector identity \( \bm{\nabla}\cdot(\phi \bm{A})= \bm{A}\cdot\left( \bm{\nabla}\phi  \right)+\phi \left( \bm{\nabla}\cdot \bm{A} \right)\), we get
\begin{align}
  \dr{}{t} \left\{ \fr{1}{V_j} \int_{\it\Omega}  {\it\Phi _j}(\bm{x}) \rho (\bm{x}, t) \mathrm{d}^3 x \right\} &= \fr{1}{V_j} \int_{\it\Omega}  \bm{J} (\bm{x}, t) \cdot \bm{\nabla} {\it\Phi _j} (\bm{x}) \mathrm{d}^3x - \fr{1}{V_j} \int_{\it\Omega}  \bm{\nabla}\cdot \left[ \bm{J} (\bm{x}, t) {\it\Phi _j}(\bm{x}) \right]\mathrm{d}^3 x  \\
  &+\fr{1}{V_j} \int_{S_n}{\it\Phi _j}(\bm{x})\rho (\bm{x}, t) \left[ \bm{u} (\bm{x})\cdot \bm{n}(\bm{x}) \right]\mathrm{d}a
  \label{eq:mmsr-1-17}
\end{align}
Applying the divergence theorem yields
\begin{align}
  \dr{}{t} \left\{ \fr{1}{V_j} \int_{\it\Omega} {\it\Phi _j} (\bm{x}) \rho (\bm{x}, t)\mathrm{d}^3 x  \right\}&= \fr{1}{V_j} \int_{\it\Omega}  \bm{J} (\bm{x}, t) \cdot \bm{\nabla} {\it\Phi _j}(\bm{x})\mathrm{d}^3 x- \fr{1}{V_j} \int_{S_j} {\it\Phi _j}(\bm{x}) \bm{J} (\bm{x}, t) \cdot \bm{n}(\bm{x}) \mathrm{d}a \\
  &+ \fr{1}{V_j} \int_{S_j} {\it\Phi _j}(\bm{x}) \rho (\bm{x}, t) \left[ \bm{u}(\bm{x})\cdot \bm{n}(\bm{x}) \right]\mathrm{d}a
  \label{eq:mmsr-1-18}
\end{align}
Finally, we get
\begin{align}
  \dfrac{\bm{\nabla} {\it\Phi _j}(\bm{x})}{V_j}&= - \dfrac{\bm{E}_j(\bm{x})}{V_j} \\
  \dr{}{t} \left\{ \fr{1}{V_j} \int_{\it\Omega} {\it\Phi_j}(\bm{x}) \rho (\bm{x}, t) \mathrm{d}^3x \right\} &= \fr{1}{V_j} \int_{\it\Omega} \bm{J} (\bm{x}, t) \cdot \bm{\nabla} {\it\Phi _j}(\bm{x}) \mathrm{d}^3x =- \fr{1}{V_j} \int_{\it\Omega} \bm{J} (\bm{x}, t) \cdot \bm{E}_j(\bm{x}) \mathrm{d}^3 x
  \label{eq:mmsr-1-19}
\end{align}
We consider that the conductor currents, $i_j(t)$, are positive when they move away from the conductor. 
Then,
\begin{equation}
  i_j(t)= - \dr{q_j(t)}{t}
  \label{eq:mmsr-1-20}
\end{equation}
The (induced) auxiliary current for conductor $j$ is defined as follows:
\begin{equation}
  I_j(t)= \fr{1}{V_j} \int_{\it\Omega} \bm{J} (\bm{x}, t) \cdot \bm{\nabla} {\it\Phi_j}(\bm{x}) \mathrm{d}^3x = - \fr{1}{V_j} \int_{\it\Omega} \bm{J} (\bm{x}, t) \cdot \bm{E}_j(\bm{x}) \mathrm{d}^3 x
  \label{eq:mmsr-1-21}
\end{equation}
If we consider a moving point charge (one type only, positive or negative), whose position is defined by the kinematic equation $\bm{x}_q= \bm{x}_q(t)$, then its velocity will be $\bm{v}_q = \dot{\bm{x}}_q$. 
In this case, the charge density and the corresponding current density can be expressed as follows:
\begin{gather}\label{eq:mmsr-1-22}
  \begin{aligned}
    \rho   & = q\delta(\bm{x} - \bm{x}_q) \\
    \bm{J} & = \rho  \bm{u}= q \bm{u}\delta(\bm{x}- \bm{x}_q)
  \end{aligned}
\end{gather}
Eq.~\eqref{eq:mmsr-1-21} yields
\begin{equation}
  I_j (t)=-q \fr{\bm{E}_j\left( \bm{x}_q(t) \right)\cdot \bm{u}\left( \bm{x}_q(t) \right)}{V_j}
  \label{eq:mmsr-1-23}
\end{equation}
The value of $\bm{E}_j$ and velocity $\bm{u}$ are refereed at time $t$ and space point $\bm{x}_q$ of charge $q$.

\noindent Eq. \eqref{eq:mmsr-1-14} becomes
\begin{align}
  I_j(t)&= i_j(t)+ \sum_{l=1}^{N} c_{jl} \dr{v_l (t)}{t} \\
  I_j(t)&= i_j(t)+ C_{jj} \dr{v(t)}{t}+ \sum_{l=1}^{N} C_{jl} \dr{\left( v_j(t)-v_l(t) \right)}{t}, \quad j=1, \ldots, N
\label{eq:mmsr-1-24}
\end{align}
This technique is called \textsl{the method of weighting field} $\bm{E}(\bm{x})/V_j=-\bm{\nabla} {\it\Phi _j}(\bm{x})/V_j$. 
It is noteworthy that $\bm{E}_j(\bm{x})$ is an auxiliary, purely electrostatic field depending solely on the conductor geometry and the dielectric between the conductors. 
It is calculated with no free moving charges in the space between the conductors, and for the respective auxiliary potentials we have $V_l \neq 0$, $V_j \neq 0$ $\forall j\neq l$. 
These potentials are not related to the biasing voltages of the detector during normal operation.

\noindent The solution to the problem, that is the determination of the various signal voltages and signal currents, is obtained by solving the system of (differential) equations shown in Eqs. \eqref{eq:mmsr-1-24}, combined with the equations that hold for the external circuit. 
The equations for the external circuit relate the currents $i_k(t)$ and voltages $v_l(t)$. 
Eqs. \eqref{eq:mmsr-1-24} show that, when charges move within the detector space, current sources $I_j(t)$ are related to the respective detector electrodes.
According to Eqs. \eqref{eq:mmsr-1-24}, we get the equivalent electrotechnical circuit of Figure \ref{fig:mmsr-1-2}, where these current sources are part of the total network, that consists of the ``internal'' network, i.e. capacities $C_{ij}$, sources $I_j$ and of the external network. 
Going backwards one could see that indeed for the equivalent circuit of the detector, the second set of the Eqs. \eqref{eq:mmsr-1-24} holds.

\noindent Included in the external circuit is the circuit providing bias for the detector. 
Usually, the bias circuit is designed in such a way that it does not affect much the signal forms in the detector outputs. 
For this reason it is often not taken into account when making calculations.

\noindent Thus, instead of solving the differential Eqs. \eqref{eq:mmsr-1-24} in combination with the external circuit equations, we can solve the electrotechnical problem depicted in Figure \ref{fig:mmsr-1-2}, where we have current sources $I_j(t)$ and the total (internal and external) network. 
We can solve the equations by means of different circuit solving methods, such as the method of analyzing in complex space, where the Laplace transform is used.

\noindent The charges move between the electrodes for finite times.
In general, after all charges have been collected the current density is $\bm{J} (\bm{x}, t) =0$. 
Signal voltages and currents are still present on the detector outputs. 
Simple cases will be presented in the following sections. 
We have to remember that (ideal) current sources have infinite resistance, so when their current is zero they represent an open circuit.

\noindent We comment on a useful result about total charge through the current source of each conductor, see book by W. Blum, W. Riegler, L. Rolandi ~\cite{blum}. 
If a point charge $q$, is moving along a trajectory $\bm{x} (t) $, from position $ \bm{x}_{0} (t_0) $ to position $ \bm{x}_{1} (t_1) $, the total charge that flows through current source number $n$, connected to conductor $n$, is given by Eq. \eqref{eq:mmsr-2-25}
\begin{gather}\label{eq:mmsr-2-25}
 \begin{aligned}
  Q_{n} = \int_{t_{0}}^{t_{1}} I_{n} \mathrm d t = - \fr{q}{V_{n}} \int_{t_{0}}^{t_{1}} \left ( \bm{E}_{n} \left( \bm{x}(t) \right) \right ) \cdot \bm{u} \left( \bm{x} (t) \right ) \mathrm{d} t = \fr{q}{V_{n}} \left ( { \it \Phi} _{n} ( \bm{x_{1}}) - { \it \Phi} _{n} ( \bm{x_{0}} ) \right )
\end{aligned}
\end{gather}

\noindent We simplified our path description by not including the index $q$. 
The charge depends only on the end points of the trajectory, it does not depend on the specific path. 
If a pair of charges $q>0$, $-q$ are at a point $ \bm{x}_{0} $, where they were produced, and after some time, charge $q$ moves and arrives at position $\bm{x}_{1} $ while $-q$ moves and reaches position $ \bm{x}_{2}$, the total charge through the current source is given by Eq. \eqref{eq:mmsr-2-26}
\begin{gather}\label{eq:mmsr-2-26}
 \begin{aligned}
 Q_{n} = \fr{q}{V_{n}} \left ( \it \Phi _{n} ( \bm{x_{1}} ) - \it \Phi _{n} ( \bm{x_{2}} ) \right )
\end{aligned}
\end{gather}
If charge $q$ moves to the surface of conductor (electrode of the detector) $n$ while charge $-q$ moves to the surface of some other electrode, the total charge through the source of the $n$ electrode is equal to $q$. 
When both charges move to other electrodes, the total charge through the $n$ source is zero. 
The conclusion is that, after all charges have arrived at the different electrodes, the total charge through the source of electrode $n$ is equal to the charge that has arrived at electrode $n$. 
From this, one also concludes that the above currents on electrodes that do not receive any charge are bipolar.

\noindent At this point we make some comments related to resistive Micromegas detectors under study to be used among other experiments in the ATLAS experiment. 
The readout strips (conductors) are placed underneath resistive strips, separated by an insulating layer. 
It is easy to examine the induced current signals in two extreme cases, a) when the resistivity of resistive strips is extremely high so they behave as pure dielectric materials, 
then the situation is simple and can be treated easily. 
The negative charges do not run all the way to the readout strips,
they stop on the resistive strips
 and at the same time the weighting field is appropriately changed. 
We assume there are ``induced'' currents only on the readout strips. 
The other extreme case b) is when resistivity is so small that resistive strips are like ideal (grounded) conductors. 
In this case there are not induced currents on the readout strips, they are shielded, and the signals on the readout strips are due to the couplings between resistive and readout strips. 
The general case is more complicated and we believe that leads to both, induced signals on the readout strips and signals due to couplings with the resistive strips. 
Probably a simpler case to be analyzed involves a resistive plane instead of resistive strips. 
For the general case one has to make use of the work of Werner Riegler, reference~\cite{riegler} and the references in that work. 

\section{Examples}

\subsection{Point charge moving within a detector of two conductors}

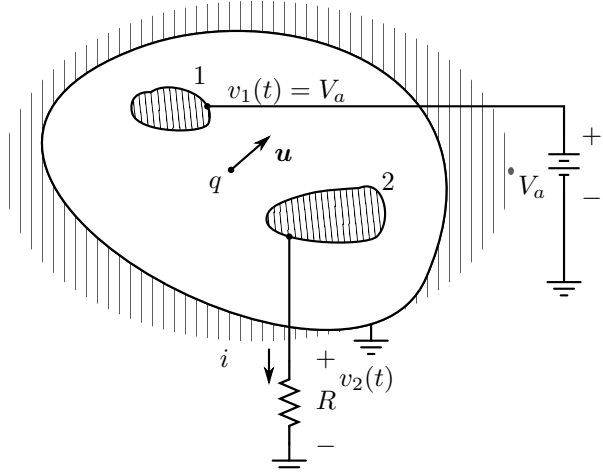
\begin{figure}[htb]
  \centering
  \input{#figures/det2cond.pdf_tex}
\caption{Detector setup with two conductors and a surrounding conductor at zero potential.}\label{fig:mmsr-p1-1}
\end{figure}

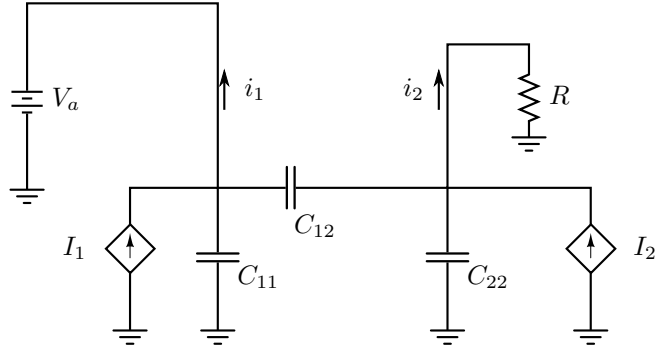
\begin{figure}[htb]
  \centering
  \input{#figures/equivdet2cond.pdf_tex}
  \caption{The equivalent electrotechnical circuit of Figure~\ref{fig:mmsr-p1-1}.}\label{fig:mmsr-p1-2}
\end{figure}

\noindent We consider a detector of two conductors, surrounded by a grounded conductor, as shown in Figure \ref{fig:mmsr-p1-1}. 
The signal is formed on conductor 2, while conductor 1 is connected to the bias source. 
The equivalent electrotechnical circuit is shown in Figure \ref{fig:mmsr-p1-2}.

\noindent
Using the first equations from Eq. \eqref{eq:mmsr-1-24}, we find
for each conductor
\begin{align}
  I_1(t) & = i_1(t)+c_{11} \dr{v_1(t)}{t} +c_{12} \dr{v_2(t)}{t} \\
         & = - \fr{1}{V_1} \int_{\it\Omega } \bm{J} (\bm{x}, t) \cdot \bm{E}_1(\bm{x}) \mathrm{d}^3 x \\
  I_2(t) & = i_2(t)+c_{12} \dr{v_1(t)}{t}+c_{22} \dr{v_2(t)}{t} \label{eq:mmsr-p1-1} \\
         & = - \fr{1}{V_2} \int_{\it\Omega} \bm{J} (\bm{x}, t) \cdot \bm{E}_2(\bm{x})\mathrm{d}^3 x
\end{align}
We have $v_l = V_a$, so $\mathrm{d}v_1/\mathrm{d}t=0$. 
For a point charge located at time $t$ located in position $\bm{x}_x(t)$ with velocity $\bm{u}(\bm{x}_q(t))$, the following equations hold for $I_1, I_2$
\begin{gather}
  \begin{aligned}
    I_1(t)&= -q \fr{\bm{E}_1\left( \bm{x}_q(t) \right)\cdot \bm{u}\left(\bm{x}_q(t)\right)}{V_1}, \qquad V_2=0  \\
    I_2(t)&= -q \fr{\bm{E}_2\left( \bm{x}_q(t) \right)\cdot \bm{u}\left(\bm{x}_q(t)\right)}{V_2}, \qquad V_1=0
  \end{aligned}
\end{gather}
$i_1(t)$ is of no interest, so out of Eqs. \eqref{eq:mmsr-p1-1} we only keep the equation of conductor 2, namely
\begin{equation}
  I_2(t)= i_2(t)+ c_{22} \dr{v_2(t)}{t}
  \label{eq:mmsr-p1-3}
\end{equation}
where $v_2(t)$ and $i_2(t)$ are the two signal forms on the detector output. 
For the resistor connected to conductor 2, we have $i_2= v_2/R$, therefore Eq. \eqref{eq:mmsr-p1-3} yields
\begin{equation}
  I_2(t)= \fr{v_2(t)}{R}+c_{22} \dr{v_2(t)}{t}
  \label{eq:mmsr-p1-4}
\end{equation}
The solution of this (differential) equation is given by \cite{rossi}
\begin{equation}
  v_2(t)= c_2 (0)+e^{-t/(Rc_{22})} \fr{1}{c_{22}} \int_0^{t} e^{t'/(Rc_{22})} I_2(t') \mathrm{d}t'
  \label{eq:mmsr-p1-5}
\end{equation}
If the two conductors of the setup in Figure \ref{fig:mmsr-p1-1} constitute an ideal capacitor, then the following equations hold
\begin{align}
  - c_{12} & = - c_{21} = c_{11} = c_{22} \\
  C_{11}   & = C_{22} = 0 \\
  C_{12}   & = C_{21} = -c_{12} = - c_{21} = c_{22} = C_{\textrm{d}}
  \label{eq:mmsr-p1-6}
\end{align}
This means that there is only one (independent) coefficient, the capacitance of the detecting conductor (electrode), with respect to the other electrode. 
The characteristic of this setup is that, when these conductors have opposite charges of the same magnitude, all the field lines beginning from one conductor end on the other. 
It should be noted that this is not true in the general case, of a pair of conductors.

\noindent We are going to examine two extreme cases that could approximately show up in practice. 
We will assume that $v=v_2$, $I=I_2$.

\noindent In the first case $C_{\textrm{d}}R\gg \tau$, where $\tau $ is the time needed for $v$ to change significantly, i.e. $C_{\textrm{d}}R |\mathrm{d}v/\mathrm{d}t| \gg |v|$; then we can omit term $v$ and we have, approximately, from Eq. \eqref{eq:mmsr-p1-4}
\begin{gather}
  \begin{aligned}\label{eq:mmsr-p1-7}
    I(t)R &= C_{\textrm{d}}R \dr{v(t)}{t}, \qquad I(t)=
    C_{\textrm{d}}\dr{v(t)}{t} \\
    \dr{v(t)}{t}&= \fr{I(t)}{C_{\textrm{d}}}, \qquad v(t)= v(0) +
    \frac{1}{C_{\textrm{d}}} \int_0^{t} I(t^{\prime}) \mathrm{d}t^{\prime}
  \end{aligned}
\end{gather}
This is equivalent to the case where the resistor ($R=\infty$) does not exist and $I$ is charging capacitance $C_{\textrm{d}}$. 
This is called the ``\textsl{voltage mode of the detector operation}''.

\noindent In the other case $C_{\textrm{d}}R\ll \tau$, then $C_{\textrm{d}}R|\mathrm{d}v/\mathrm{d}t|\ll |v|$; therefore, we have approximately
\begin{equation}\label{eq:mmsr-p1-8}
  IR=v 
\end{equation}
This is equivalent to a circuit, where there is no capacitor, $C_{\textrm{d}}=0$ and the whole current $I$ is passing through the resistor. 
This is called the ``\textsl{current mode of the detector operation}''.

\noindent We can solve the problem by considering the electrotechnical circuit of Figure \ref{fig:mmsr-p1-2}. 
Indeed, by applying Kirchhoff's laws we finally end up with (differential) Eq. \eqref{eq:mmsr-p1-4}. 
Remember that it can be shown that $c_{22} = C_{22}+C_{12}$.

\subsection{Cylindrical detector with circular cross-section and a circular wire conductor on its axis, MDT like detector} \label{subsec:ex2}

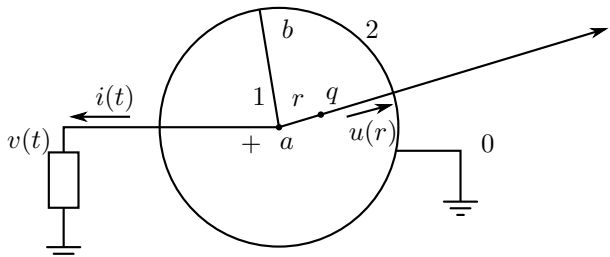
\begin{figure}[htb]
  \centering
  \input{#figures/crosssectcond.pdf_tex}
  \caption{A cylindrical detector with circular cross section.}\label{fig:mmsr-p1-3}
\end{figure}

\begin{figure}[htb]
  \centering
  \input{#figures/equivcrosssectcond.pdf_tex}
  \caption{The equivalent electrotechnical circuit of Figure~\ref{fig:mmsr-p1-3}.}\label{fig:mmsr-p1-4}
\end{figure}
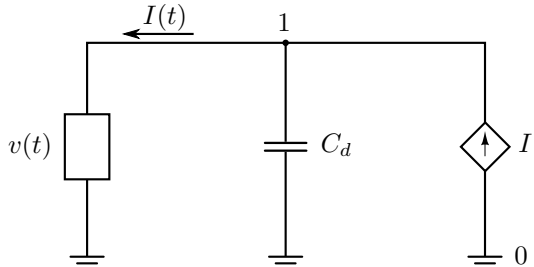

\noindent For this analysis one could use references \cite{blum} and \cite{rossi}.
A cross-section of a singe MDT detector of the ATLAS experiment ~\cite{atlastdr} is illustrated in Figure \ref{fig:mmsr-p1-3}. 
The equivalent electrotechnical circuit can be seen in Figure \ref{fig:mmsr-p1-4}.

\noindent We assume that the detector is a cylinder of much bigger length than its cross sectional diameter. 
We can thus disregard the fringe field effects at the two ends of the detector, i.e. practically, the field is radial everywhere inside the detector.
Within the detector a point charge $q$ is moving radially from the center outwards. 
In Figures \ref{fig:mmsr-p1-3} \& \ref{fig:mmsr-p1-4} the positive bias voltage, $V_a$, of the central detector electrode, conductor 1 with respect to enclosure 0, which coincides with conductor 2 (potential zero) is not shown. 
We assume that the circuit is such that the bias circuit does not affect the detector signal.

\noindent According to the above analysis, we examine the detector as an one-conductor device with $N=1$. 
It can be easily understood that the only capacitance coefficient involved is the detector capacitance, $C_{\textrm{d}}$, which is the capacitance formed by the two conductors, given by the following formula
\begin{equation}
  C_{\textrm{d}}= C_l l = l \fr{2\pi \varepsilon }{\ln(b/a)}
  \label{eq:mmsr-p1-9}
\end{equation}
where $l$ is the detector length and $C_l$ is its capacitance per unit length.

\noindent The bias potential biases conductor 1 positively with respect to the enclosure, 0. 
The (inner) radius of the outside conductor is $b$ and the radius of the central conductor is $a<b$. 
The auxiliary electric field in the space of the cylindrical detector for potential $V$ in electrode 1 (i.e. the central electrode of the cylindrical detector) is radial with an outwards direction, and is given by
\begin{equation}
  E(\bm{x}) \bm{{e}}_r = \fr{C_l V}{2\pi \varepsilon } \fr{1}{r} \bm{{e}}_r = \fr{V}{\ln (b/a)} \fr{1}{r} \bm{{e}}_r
  \label{eq:mmsr-p1-10}
\end{equation}
We use cylindrical coordinates, as usual. 
It has been taken into account that the capacitance per unit length is given by
\begin{equation}
  C_l = \fr{2\pi \varepsilon }{\ln (b/a)}
  \label{eq:mmsr-p1-11}
\end{equation}
For gaseous detectors, it holds that $\varepsilon _{\mathrm{r}} \approx 1$, $\varepsilon = \varepsilon_\mathrm{r} \varepsilon _0 \approx \varepsilon _0$.

\noindent From Eq. \eqref{eq:mmsr-1-23} we get
\begin{equation}
 I(t)=-q \fr{1}{\ln(b/a)} \fr{1}{r(t)} \bm{u}(\bm{x})\cdot \bm{{e}}_r
  \label{eq:mmsr-p1-12}
\end{equation}
The motion of the point charge is radial, so we have
\begin{equation}
 I(t)= - q\fr{1}{\ln(b/a)} \fr{1}{r(t)}u\left( r(t) \right)
  \label{eq:mmsr-p1-13}
\end{equation}
$u\left( r(t) \right)$ is positive only if the motion is from the center outwards and negative in the opposite case.

\noindent We accept that a point charge, $q$, started at time $t=0$ from position $r=r_0>a$, where $a$ is the wire radius and is moving with instantaneous velocity $u\left( r(t) \right)$. 
This start position is, in practice, found to be very close to the wire surface. 
This is because ionization occurs very close to the wire.  
Positive ions move inside the detector towards the cathode, electrode 2, and negative particles (electrons) move towards the anode, electrode 1.

\noindent Since electrons are collected in a very short time one could ignore their contribution to the detector signal. 
Thus, we examine only the positive ions, assuming that there is only one kind. 
For ions, the drift velocity (for a wide range of fields) is given by
\begin{equation}
  u= \mu  E_a
  \label{eq:mmsr-p1-14}
\end{equation}
where $ \mu $ (mobility) is constant. 
It must be noted that field $E_a$ is the real field in the detector due only to the high-voltage detector bias, ignoring the effect of space charges and assuming that the voltage drop in the detector connected to the high-voltage supply is negligible compared to the high voltage. 
Using Eqs. \eqref{eq:mmsr-p1-10} and \eqref{eq:mmsr-p1-14}, we get
\begin{equation}
  u= \mu \fr{V_a}{\ln(b/a)}\fr{1}{r}
  \label{eq:mmsr-p1-15}
\end{equation}
It is evident that positive point charge $q$ will be moving from the anode (central electrode, signal electrode) towards the cathode (external electrode) and we will have
\begin{equation}
  I= - \fr{q\mu V_a}{ \ln^2(b/a) } \fr{1}{r^2}
  \label{eq:mmsr-p1-16}
\end{equation}
The equivalent circuit contains a current source that is connected to the external network shown in Figure \ref{fig:mmsr-p1-4}. 
For the point charge of ions we need its position as a function of time.
From the equations above we get
\begin{align}
  \dr{r}{t} & = \mu \fr{V_a}{ \ln(b/a)} \fr{1}{r}, \qquad r \mathrm {d}r= \mu \fr{V_a}{ \ln(b/a)} \mathrm {d}t, \qquad \int_{r_0}^{r}r^{\prime} \mathrm{d}r^{\prime}= \int_0^{t}\mu  \fr{V_a}{\ln(b/a)} \mathrm{d}t^{ \prime} \\
  r^2       & = r_0^2+2 \mu \fr{V_a}{ \ln(b/a)}t, \qquad r^2=r_0^2 \left( 1+ \fr{t}{t_0} \right) \\
  t_0       & = \fr{r_0^2}{2 \mu V_a} \ln(b/a) \label{eq:mmsr-p1-17} \\
  T         & = \fr{(b^2-r_0^2) \ln(b/a)}{2\mu V_a} \\
  I(t)      & =- \fr{q\mu V_a}{r_0^2 \ln^2 (b/a)} \fr{1}{(1+t/t_0)}
\end{align}
The $t_0$ parameter defines the time scale for the charge carrier motion and the induced signal and $T$ is the arrival time of point charge on the external enclosure.

\noindent From Figure \ref{fig:mmsr-p1-4} and with the help of Eqs. \eqref{eq:mmsr-1-24} we get
\begin{gather}\label{eq:mmsr-p1-18}
  \begin{aligned}
    I(t)  & = \fr{v(t)}{R}+ C_{\textrm{d}}\dr{v(t)}{t}\\
    I(t)R & = v(t)+RC_{\textrm{d}} \dr{v(t)}{t}
  \end{aligned}
\end{gather}
We will analyze two cases:

\noindent {\bf (a)} Assume $RC_{\textrm{d}}\ll \tau $.

\noindent Here, the characteristic time is $ \tau =t_{0} $. 
Characteristic time is, in general, different for the various carrier types, ions and electrons. 
We omit the term containing $RC_{ \textrm{d}} $ and we therefore have approximately
\begin{equation}
  v(t)= I_{q}R \quad \mathrm{for} \quad 0<t\leq T \qquad \mathrm{and} \qquad
  v(t)=0 \quad \mathrm{for} \quad t>T
  \label{eq:mmsr-p1-19}
\end{equation}
We might say that with this approximation, the signal lasts for as long as the duration of the charge motion in the detector space (time $T$), since capacitive effects have been omitted. 

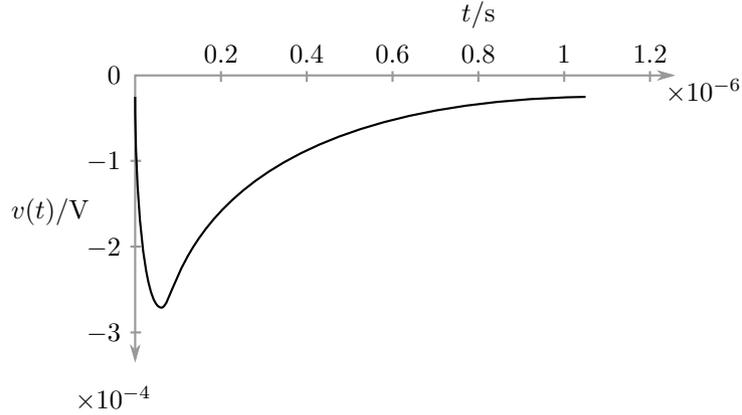
\begin{figure}[htbp]
  \centering
  \input{#figures/indicativesignal.pdf_tex}
  \caption{Indicative signal $v(t)$ as a function of time $t $ for a cylindrical detector.}\label{fig:mmsr-1-5}
\end{figure}

\noindent
We will determine the charge induced on the detector during the motion of positive charge $q$. 
We define this charge as positive if we have positive charge moving towards the examined electrode. 
To find the charge induced from time $t=0$ to time $t\leq T$, we integrate the current with respect to time. 
Therefore, we get
\begin{align}
  Q(t) & = \int_0^{t} I(t^{\prime})\mathrm{d}t^{\prime}= \fr{q}{ \ln(b/a)} \int_0^{t} \fr{u}{r}\mathrm{d}t^{\prime}= \fr{q}{\ln(b/a)} \int_{r_0}^{r}\fr{u}{r} \fr{\mathrm{d}r}{u}= \fr{q}{\ln(b/a)}\ln \left( \fr{r}{r_0} \right) \\
       & = \fr{q}{2 \ln(b/a)} \ln\left( 1+\fr{t}{t_0} \right)>0 \label{eq:mmsr-p1-21} \\
       \mathrm{and\  total\  charge}\\
  Q_\textrm{T}  & = \fr{q}{\ln(b/a)}\ln \left( \fr{b}{r_0} \right) >0
\end{align}
{ \bf (b)} The other extreme case is for $RC_{ \textrm{d}} \gg \tau =t_0 $. 
In this case, it is as if there were no resistor, $R=\infty$, in the equivalent circuit. 
In Eqs. \eqref{eq:mmsr-p1-18} we omit the $v(t)$ term, therefore
\begin{equation}
  I(t)= C_{\textrm{d}} \dr{v(t)}{t}
  \label{eq:mmsr-p1-22}
\end{equation}
The solution of the above equation, after using Eq. \eqref{eq:mmsr-p1-17}, yields
\begin{equation}
  v(t)= v(0)+\fr{1}{C_{\textrm{d}}} \int_0^{t}I(t^{\prime})\mathrm{d}t^{\prime}= - \frac{1}{C_{\textrm{d}}} \int_0^{t} I(t^{\prime})\mathrm{d}t^{\prime} = - \fr{q}{2C_{\textrm{d}} \ln(b/a)}\ln\left( 1 +\fr{t}{t_0} \right), \qquad 0\leq t \leq T
  \label{eq:mmsr-p1-23}
\end{equation}
and
$$
v(t) =
 v(T)  \qquad  \mathrm{for} \quad t>T
\label{eq:mmsr-p1-23b}
$$
In this case, after time $T$, the capacitor-detector remains charged at voltage $v(T)$.
In the general case, with no approximations, we give the solution of Eq. \eqref{eq:mmsr-p1-18} for the above cylindrical detector case.
The result, with the initial condition $v(0)=0$ and with $RC_{ \textrm{d}}=\tau_{ \textrm{d}}$, yields
$$
v(t) =
\begin{cases}
     - \fr{q\mu V_aR}{r_0^2 \ln^2(b/a)} \fr{t_0}{ \tau_{ \textrm{d}}} e^{{-(t+t_0)}/{ \tau _{\textrm{d}}} } \left[ E_1 \left( -\fr{t_0}{ \tau _{ \textrm{d}}} \right)- E_1 \left( - \fr{t+t_0}{\tau _{ \textrm{d}}}\right) \right] & \qquad \mathrm{for} \quad 0\leq t \leq T, \\
      &  \\
    v(T) e^{ {-(t-T)}/{ \tau _{ \textrm{d}}} } & \qquad \mathrm{for} \quad t>T
\end{cases}
$$
where $E_1(x) = \int_x^{ \infty} e^{-t}/t \, \mathrm{d}t $ is the exponential integral. 
In Figure \ref{fig:mmsr-1-5} we see an indicative signal (potential) waveform of a cylindrical detector, $v(t)$, which corresponds to a positive point charge (ions) of one type moving radially, before being collected. After time $T$, capacitor $C_\textrm{d}$ 
discharges through resistor $R$.

\subsection{Signals in a Micromegas detector}\label{subsec:mm}

\subsubsection{Overview}

\noindent We stress that the resistive strips have very large resistivity so they play the role of insulating material. 
We ignore signals in the resistive strips and couplings to the readout strips.

\noindent The geometry of the detector is a plane geometry, see Figure \ref{fig:ionesignal1} (see ~\cite{FirstReference, SecondReference}). 
We imagine two Cartesian coordinates, one, $z$, is normal to the mesh and strip plane, and the other, $x$, is along the above planes normal to the $z$. 
We ignore the end effects that involve a very small area of the whole detector. 
Furthermore, the detector has small dimensions, so electromagnetic wave propagation is ignored.

\noindent We assume there is not longitudinal or transverse diffusion of charges. 
This means that a point charge remains a point charge as it moves. 
Let a point charge $q$ at a point A inside the gas multiplication region moving with a known drift velocity $ \bm{u}$. 
The biasing potentials determine the motion of charges in the detector space. 
We consider that the biasing potentials are dc (no time dependence) so the drift velocity depends only on the position of the moving point charge. 
To calculate the signal on a readout strip electrode, we use the generalized Shockley-Ramo theorem, proved above. 
We have to calculate the current sources corresponding to each one readout strip for each type of moving charge. 
To do so, we put one at a time readout strip (electrode), $a$, at positive potential $V_a$ and all other electrodes at zero potential. 
The electric field at point A, position vector $ \bm{x}$, is $ \bm{E}_A$. 
The point A is the position of the moving point charge. 
This is an auxiliary field that has no relation to the actual biasing (dc) potentials of the electrodes of the ``working'' detector. 
We stress that, in practice, the biasing potentials are much bigger than the signal potentials.

\noindent The ratio (weighting field) $\bm{E}_ \textrm{A}/V_a$ depends only on the geometry and the electric properties of the materials in the space inside the detector. 
It is clear that this ratio depends on the position $ \bm{x}$ of the point A. 
It has dimension of inverse length, so in the SI is measured in $\mathrm{m}^{-1}$.

\noindent The source current which is equal to the auxiliary (induced) current on electrode $a$ due to the motion of point charge $q$ at A, is given by
\begin{equation}
  I_a = -q \fr{\bm{E}_A\cdot \bm{u}}{V_a}
  \label{eq:ionesignal1}
\end{equation}
In a Micromegas detector we have both electrons and positive ions drifting in opposite directions, each time starting from the same point in space. 
The electrons move with much higher speeds than the (positive) ions and we treat each sign of charge separately. 
The biasing is dc and such that the electrons move from the mesh towards the resistive strips, where they are absorbed and stop.

\noindent For a varying continuous charge distribution with current density $ \bm{J} (\bm{x}, t)$, the auxiliary current is
\begin{equation}
  I_a(t)= - \fr{1}{V_a} \int_\Omega  \bm{J} (\bm{x}, t)\cdot \bm{E}(\bm{x})\, \mathrm{d}^3 x
  \label{eq:ionesignal2}
\end{equation}
The integral is over the volume where the charge is distributed.

\noindent Figure \ref{fig:mmsr-1-2} shows the equivalent circuit of a multi-electrode (three electrodes) detector connected to an external circuit. 
To determine the weighting field, one has to solve the Laplace equation assuming that the particular strip has a potential $V_a>0$ and all other strips and the mesh are grounded. 
One determines the potential $\it\Phi (x,z)$ of such a plane geometry and then from that the weighting field is determined. 
The potential and the field is a complicated function of position. 
There are simplified geometries for which the field can be easily calculated.  
Summarizing: we have to calculate the signals induced on the strips of the detector, when a specific point charge moves in a known way inside the amplification region of the detector. 
To do so, for all readout strips we calculate the equivalent current sources. 
At the end, we use the equivalent circuit shown in Figure \ref{fig:mmsr-1-2} and the detector signals are determined.

\subsubsection{Signal formation in resistive Micromegas}

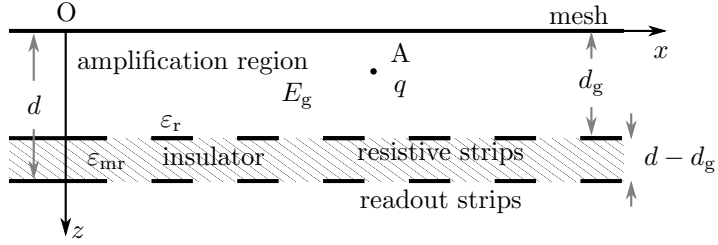
\begin{figure}[htb]
  \centering
  \input{#figures/micromegas_geo.pdf_tex}
  \caption{Micromegas geometry of the amplification region.}\label{fig:ionesignal1}
\end{figure}

\noindent The geometry of the detector is shown in Figure \ref{fig:ionesignal1} (see ~\cite{SecondReference}). 
When electrons enter the space between MESH and resistive strips (gas amplification region), they produce pairs of ions and electrons forming showers.
The charge is multiplied
 as the electrons move from the MESH towards to the resistive strips. 
The multiplication of charge depends on the (first) Townsend coefficient $\alpha$ which in general, among other things, is a function of the electric field. 
If $n$ is the number of one type of charged particles at a certain point, then at a nearby point along the path of the moving charge the increase of the charges will be $ \textrm{d}n=\alpha n \textrm{d}r$, where $ \textrm{d}r$ is the (infinitesimal) distance between the two space points. 
In order to calculate the increase of charge one has to integrate the above formula. 
We will simplify our calculations by assuming that the Townsend coefficient is constant, independent of the electric field. 

\noindent
We assume that the resistive strips have very high resistance so they behave like insulating material that has the same effect as the shaded insulator between resistive and readout strips, shown in Figure \ref{fig:ionesignal1}.

\noindent We assume that the resistive strips have the same relative permittivity, $\varepsilon _{ \textrm{mr}}$ with the material between the two types of strips.

\noindent Let us consider a point charge $q$ at a point A inside the gas multiplication region moving with velocity $ \bm{u}$.
The auxiliary (induced) current on electrode $a$ due to the motion of this point charge $q$, is given by the Eq. \eqref{eq:ionesignal1}.

\noindent A simple case is the calculation of the signal on the mesh which is one very large, practically plane, electrode and the weighting field for this electrode is the field formed inside a parallel plate capacitor, assuming  that the field even near the strips is homogeneous. 
We will consider that the point charge is negative and its motion is normal to the strip planes. 
In the case, of the almost ideal plane capacitor geometry, it is easy to show that (see Figure \ref{fig:ionesignal1})
\begin{equation}
  \fr{E_a}{V_a}= \fr{1}{d}  \left[ \fr{d_\textrm{g}}{d}\left( 1- \dfrac{\varepsilon
  _\textrm{r}}{\varepsilon_{\textrm{mr}}} \right)+\fr{\varepsilon _\textrm{r}}{\varepsilon _{\textrm{mr}}} \right]^{-1}
  \label{eq:ionesignal3}
\end{equation}
This quantity, as a vector, is directed from the mesh to the strips.  
Usually, we define the effective gap $d_{ \textrm{eff}}$ as follows:
\begin{equation}
  d_{\textrm{eff}}= d
\left[ \fr{d_\textrm{g}}{d}\left( 1- \dfrac{\varepsilon
  _\textrm{r}}{\varepsilon_{\textrm{mr}}} \right)+\fr{\varepsilon _\textrm{r}}{\varepsilon _{\textrm{mr}}} \right]  
  \label{eq:ionesignal4}
\end{equation}
so the relation of Eq. \eqref{eq:ionesignal3} is reduced to
\begin{equation}
  \fr{E_a}{V_a}= \fr{1}{d_{\textrm{eff}}}
  \label{eq:ionesignal5}
\end{equation}
For the moving point charge $q<0$ with velocity $u$, which has the same direction with the auxiliary field, the auxiliary current is
\begin{equation}
  I_{ \textrm{mesh}}= - \fr{qu}{d_{ \textrm{eff}}}
  \label{eq:ionesignal6}
\end{equation}
For the strips in general, things are more complicated because the weighting field for each strip is not so homogeneous, it varies a lot from point to point.

\begin{figure}[htb]
  \centering
  \input{#figures/chdistr_shower.pdf_tex}
  \caption{``Static'' ion charge distribution due to shower development in the gas amplification gap.}\label{fig:ionesignal3}
\end{figure}
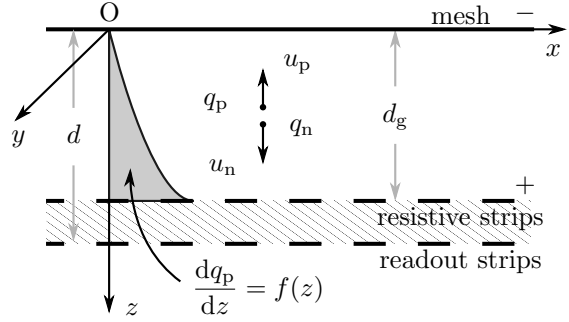

\noindent The charge multiplies starting from the point of entrance of the electron point charge, entering from the mesh into the multiplication region.
We assume the charges do not diffuse. 
The negative charge remains a variable value point charge, moving towards the resistive strips.
The speed of the negative point charge is \(u_{\text{n}}\ (>0)\) the same at every point.
The negative charge moves till it arrives at the resistive strip plane at time \(T_{\text{n}} = d_\textrm{g}/u_{\text{n}}\). 
After this time, the multiplication of charges stops.
The auxiliary current of the negative charge becomes zero.
Things are more complicated for the positive ion charges.
The positive charges increase with the same rate as the negative charges, but they move much slower in the opposite direction, towards the mesh, all with velocity \( u_{\text{p}} \, (>0)\). 
This means that they are not a point charge; instead, they are distributed inside the amplification region from point \(z=0\) (see Fig. \ref{fig:ionesignal3}) to the point where, each time, the negative charge has reached.
As time progresses (times are measured from the moment the negative point charge enters the amplification region), part of the positive charge is ``absorbed'' by the mesh.
As the distribution moves, it gets ``distorted''.
After time \(t=T_{\text{n}} = d_\textrm{g}/u_{\text{n}}\), there is no multiplication.
At this time, the existing whole ion distribution starts shifting to the mesh without ``distortion'' with velocity \(u_{\text{p}}\).
The change that reaches the mesh gets absorbed by the mesh and this lasts till all the positive charges reach this plane, at time

\[
T_{\text{p}} = \dfrac{d_\textrm{g}}{u_{\text{n}}} + \dfrac{d_\textrm{g}}{u_{\text{p}}}
\]
The ion auxiliary current lasts from \(t=0\) to \(t= T_{\text{p}}\).
We will calculate separately the auxiliary currents due to the electrons and the positive ions.
From \(\mathrm{d} n = n \alpha \mathrm{d} z\), we get \(n = n_0 e^{\alpha z}\).
Let \(e>0\) be the elementary charge (charge of the proton), then the absolute value of negative charge in the interval from \(z=0\) to \(z\) is \(q_{\text{n}}= ne= n_0 e^{\alpha z}= q_0 e^{\alpha z}\), where \(q_0\) is the absolute value of the initial negative point charge entering the amplification region from the mesh,
usually $n_0 = 1$ and $q_0=e$.
For the corresponding positive charge, we have 
\[
q_{\text{p}} = q_0 e^{\alpha z}-q_0= q_0 \left(e^{\alpha z}-1\right)
\]
All negative charge is at the point \(z\) (it is a point charge of value \(q_{\text{n}}= q_0 e^{\alpha z}\)).
It moves with velocity \(u_{\text{n}}\) towards the resistive strips. 
The auxiliary electron current for an electrode $a$ at time \(t\) is 
\[
I_{\text{n}a}(t) = \pm q_0 e^{a u_{\text{n}} t} \dfrac{E_z(z) u_{\text{n}}}{V_a}, \ \ z = u_{\text{n}} t
\]
We use the formula for a point charge, Eq. \eqref{eq:ionesignal1}.
The ``+'' sign holds for the mesh and the ``-'' for the readout strips.
This is not zero for \(0 \leq t \leq T_{\text{n}} = d_\textrm{g}/u_{\text{n}}\).
For \(t>T_{\text{n}}\), \(I_{\text{n}a}(t)=0\).
If we assume that for the mesh \(E_z = E = \text{homogeneous}\), then 
\[
\dfrac{E_z}{V_a}= \dfrac{1}{d_{\text{eff}}}
\]
So we get the previous result
\[
I_{\text{mesh}} = \dfrac{q_\text{n} u_{\text{n}}}{d_{\text{eff}}}
\]
Remember that now $q_\text{n}>0$.

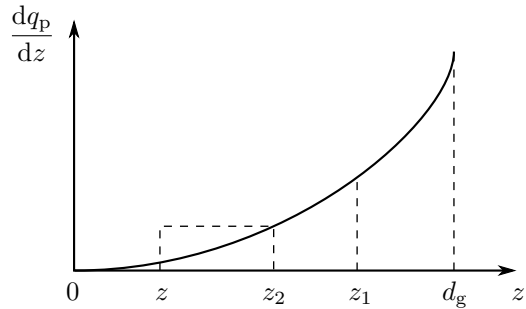
\begin{figure}[htb]
  \centering
  \input{#figures/graph10.pdf_tex}
  \caption{Charge positive ion distribution developed from a typical avalance developed in a typical gaseous detector.}\label{fig:graph10}
\end{figure}

\noindent
Let us now work with the positive ion auxiliary current.
Figure \ref{fig:ionesignal3} shows the ``static'' distribution of positive ions.
It is the distribution of ions inside the amplification region, if they were not moving.
It is obtained by differentiation of \(q_{\text{p}} = q_0 (e^{\alpha z}-1)\).
We have
\[
\drac{q_{\text{p}}}{z} = q_0 \alpha e^{\alpha z}
\]
The elementary infinitesimal charge in the range \(z\), \(z+\mathrm{d}z\) is \(\mathrm{d}q_{\text{p}} = q_0\alpha e^{\alpha z}\mathrm{d}z\).
It is useful to work with Figure \ref{fig:graph10}.
In this figure we see the static distribution from \(z=0\) to \(z=d_\textrm{g}\).
Let us assume that at time \(t_1=t\), the negative point charge reached point \(z_1 = u_{\text{n}} t_1= u_{\text{n}} t\).

\noindent At this time \(t_1=t\), the positive charge that was produced at point \(z_2\) at time \(t_2\), \( (z_2 = u_{\text{n}} t_2) \), has reached position \(z=z(t)\) at time \(t_1\). 
We have 
\begin{align}
  z(t) & = u_{\text{n}} t_2- u_{\text{p}} (t_1-t_2)\\
       & = z_2- u_{\text{p}} \dfrac{z_1}{u_{\text{n}}}+ u_{\text{p}} \dfrac{z_2}{u_{\text{n}}}
\end{align}
so 
\[
z = z_2 -\dfrac{u_{\text{p}}}{u_{\text{n}}}(z_1-z_2)
\]
and
\[
z_2= \dfrac{u_{\text{n}}}{u_{n}+u_{\text{p}}}z+ \dfrac{u_{\text{p}}}{u_{\text{n}}+u_{\text{p}}}z_1
\]
The maximum value of \(z_1=d_\textrm{g}\) and it is obtained when the negative charge reaches the resistive planes at time \(T_{\text{n}}= d_\textrm{g}/u_{\text{n}}\).
This is a very small time in comparison with the maximum drift time of motion, $T_\textrm{p}$,  of the positive charges
\[
T_{\text{p}} = \dfrac{d_\textrm{g}}{u_{\text{n}}}+\dfrac{d_\textrm{g}}{u_{\text{p}}}= T_{\text{n}} +\dfrac{d_\textrm{g}}{u_{\text{p}}}
\]
It is clear that 
\[
\drac{z}{z_2}= 1+\dfrac{u_{\text{p}}}{u_{\text{n}}}= \dfrac{u_{\text{n}}+u_{\text{p}}}{u_{\text{n}}}
\]
and
\[
\drac{q_{\text{p}}(z)}{z} = \drac{q_{\text{p}}}{z_2} \drac{z_2}{z}= \drac{q_{\text{p}}}{z_2} \dfrac{u_{\text{n}}}{u_{\text{n}}+u_{\text{p}}} = \drac{u_{\text{p}}}{z_2} \beta, \quad 0< \beta = \dfrac{u_{\text{n}}}{u_{n}+u_{\text{p}}} < 1
\]
The ``distorted'' ion distribution gives for the infinitesimal charge,
\[
\mathrm{d}q_{\text{p}} (z)= q_0 \alpha \beta e^{\alpha (1-\beta )z_1} e^{\alpha \beta z}\mathrm{d}z, \quad z_1 = u_{\text{n}}t
\]
This is the distribution when the negative charge is at point \(z_1\), \(0\leq z_1\leq d_\textrm{g}\).
It is evident that this can be used for \(0\leq t \leq d_\textrm{g}/u_{\text{n}}\).
The corresponding infinitesimal auxiliary current at electrode \(a\) is [from Eq. \eqref{eq:ionesignal1}], 
\[
\mathrm{d}I_{\text{p}a}= \pm \dfrac{E_z(z)u_{\text{p}}}{V_a} \mathrm{d}q_{\text{p}}
\]
The ``+'' sign holds for the mesh and the ``-'' sign for the readout strips.
We integrate to find the total \(I_{\text{p}a}\),
\[
I_{\text{p}a}(t)= \pm \alpha \beta  \dfrac{u_{\text{p}}}{V_a} q_0 e^{\alpha (1-\beta )z_1} \int_0^{z_1} E_z(z)e^{\alpha \beta z}\mathrm{d}z, \quad z_1= u_{\text{n}} t_1=u_{\text{n}} t
\]
This applies for \(0\leq t \leq d_\textrm{g}/u_{\text{n}}\). 
If \(E(z)\) is homogeneous, this is a reasonable assumption for the mesh, then
$E_z=E$, and 
\begin{align}
I_{\text{pmesh}}(t) & = \alpha \beta  \dfrac{u_{\text{p}}}{V_a} q_0e^{\alpha (1-\beta )z_1} E\int_0^{z_1} e^{\alpha \beta z} \mathrm{d}z \\
                    & = \dfrac{u_{\text{p}}}{V_a} q_0 Ee^{\alpha z_1} \left[1-e^{-\alpha \beta z_1}\right]
\end{align}
We have shown that 
\[
\dfrac{E}{V_a} = \dfrac{1}{d_{\text{eff}}}
\]
so 
\[
I_{\text{pmesh}} (t) = u_{\text{p}} q_0 \dfrac{1}{d_{\text{eff}}} e^{\alpha z_1} \left[1-e^{-\alpha \beta z_1}\right]
\]
In practice, \(u_{\text{p}}\) is about \(u_{\text{n}}/100\) to \(u_{\text{n}}/1000\), that means \(u_{\text{p}}\ll u_{\text{n}}\), so \(\beta \approx 1-u_{\text{p}}/n_\textrm{n}\), so 
\[
I_{\text{pmesh}}(t) \approx \dfrac{q_0}{d_{\text{eff}}} u_{\text{p}} \left[ e^{\alpha u_{\text{n}}t}- e^{\alpha u_{\text{p}}t}\right]
\]
for \(0\leq t\leq d_\textrm{g}/u_{\text{n}}\).
For times \(t> d_\textrm{g}/u_{\text{n}}\), the whole (distorted) distribution obtained when \(z=d_\textrm{g}\) drifts undistorted towards the mesh.
The distribution for \(z_1=d_\textrm{g}\) is given by 
\[
\mathrm{d}q_{\text{p}} = q_0\alpha \beta e^{\alpha \beta z} e^{\alpha (1-\beta )d_\textrm{g}}\mathrm{d}z
\]
The moving undistorted distribution is obtained by the replacement \(z\to z+(d_\textrm{g}-z_1)\).
Now \(z_1\) is the position where the maximum of the distribution occurs.
\[
\mathrm{d}q_{\text{p}} = q_0\alpha \beta e^{\alpha \beta [z+(d_\textrm{g}-z_1)]} e^{\alpha (1-\beta )d_\textrm{g}}\mathrm{d}z
\]
or finally
\[
\mathrm{d}q_{\text{p}} = q_0\alpha \beta  e^{\alpha (d_\textrm{g}-\beta z_1)}e^{\alpha \beta z} \mathrm{d}z
\]
\begin{align}
z_1 &= d_\textrm{g} -u_{\text{p}} (t-T_{\text{n}})\\ 
&= \dfrac{d_\textrm{g}}{\beta }-u_{\text{p}} t
\end{align}
We have 
$
d_\textrm{g} -\beta z_1= d_\textrm{g}-d_\textrm{g}+\beta  u_{\text{p}} t = \beta  u_{\text{p}} t , 
$
so 
\[
\mathrm{d}q_{\text{p}} = q_0\alpha \beta e^{\alpha \beta u_{\text{p}} t} e^{\alpha \beta t} \mathrm{d}z
\]
\[
\mathrm{d}q_{\text{p}} = q_0\alpha \beta e^{\alpha (d_\textrm{g} -\beta z_1)} e^{\alpha \beta z} \mathrm{d}z
\]
The induced current is
\[
\mathrm{d}I_{\text{p}a}= \pm \dfrac{E_z(z) u_{\textrm{p}}}{V_a} \mathrm{d}q_{\text{p}}
\]
The ``+'' sign is for the mesh and the ``-'' sign for the reading strips.
Integration for \(0\leq z \leq z_1\) gives
\[
I_{\text{p}a}= \pm \dfrac{q_0\alpha \beta e^{\alpha (d_\textrm{g}-\beta z_1)}u_{\textrm{p}}}{V_a} \int_0^{z_1} E_z(z) e^{\alpha \beta z}\mathrm{d}z
\]
We have \(z_1= d_\textrm{g}-u_{\text{p}}(t-T_{\text{n}})\), so finally
$
z_1= {d_\textrm{g}}/{\beta }-u_{\text{p}} t
$
since $\beta = u_n / (u_n + u_p)$  or
\[
d_\textrm{g}-\beta z_1= \beta  u_{\text{p}} t
\]
The current $I_{\text{p}a}(t)$ will be given by
\[
I_{\text{p}a}(t)=\pm \dfrac{q_0\alpha \beta e^{\alpha \beta u_{\text{p}} t}u_{\textrm{p}}}{V_a} \int_0^{d_\textrm{g}/\beta - u_{\text{p}} t} E_z(z) e^{\alpha \beta z}\mathrm{d}z, \qquad \dfrac{d_\textrm{g}}{u_{\text{n}}}+\dfrac{d_\textrm{g}}{u_{\text{p}}}>t>\dfrac{d_\textrm{g}}{u_{\text{n}}} 
\]
or \(T_{\text{p}}>t >T_{\text{n}} \).

\noindent Let us assume that we refer to the mesh when the approximation \(E=\text{homogeneous} = E_z\) is reasonable, then \(E/V_a = 1/d_{\text{eff}}\).
\[
I_{\text{pmesh}}(t)= \dfrac{q_0u_{\textrm{p}}}{d_{\text{eff}}} \left[e^{\alpha d_\textrm{g}}-e^{\alpha \beta u_{\text{p}} t} \right]
\]
Since in practice \(u_{\text{p}} \ll u_{\text{n}}\), then \(\beta \approx 1-u_{\text{p}}/u_{\text{n}}\). 
Ignoring higher order terms we get
\[
I_{\text{pmesh}}(t) \approx \dfrac{q_0u_{\textrm{p}}}{d_{\text{eff}}} \left[e^{\alpha d_\textrm{g}}- e^{\alpha \beta u_{\text{p}} t}\right]
\]
It is obvious that \(I_{\text{p}a}=0\) for \(t> \dfrac{d_\textrm{g}}{u_{\text{n}}}+\dfrac{d_\textrm{g}}{u_{\text{p}}}= T_{\text{p}}\). 



\noindent We mention that the auxiliary currents are defined with different formulae for various time intervals. 
In the general case of capacitors, inductors and so on, one should remember that the signals exist even when the auxiliary currents become zero. 
It is clear that the electron auxiliary currents last for time periods much smaller that the ion currents, $T_\textrm{n} \ll T_\textrm{p}$.

\noindent  In "classical Micromegas" there is no resistive material, the electrons reach the readout strips. It is easy to get the simpler formulae for this case, starting 
from the above formulae for resistive Mcromegas.  

\begin{figure}[htb]
  \centering
\includegraphics[scale=0.20,angle=0,width=\columnwidth]{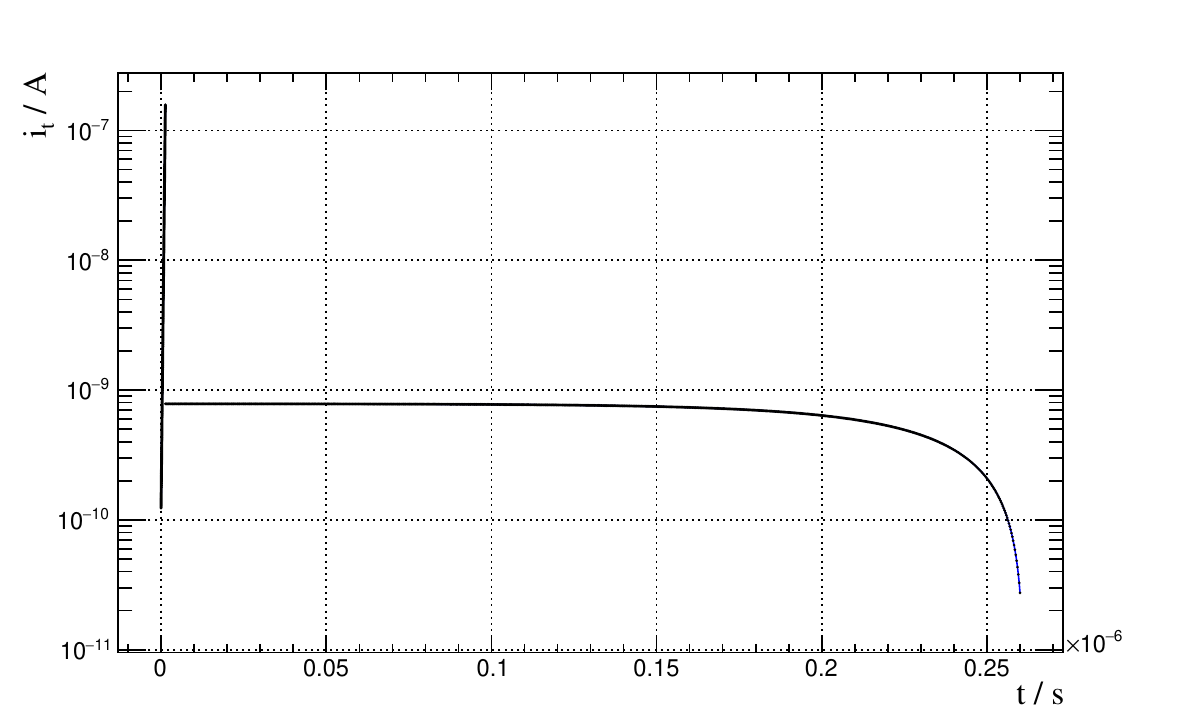}
  \caption{Total auxiliary current as a function of time. 
The peak at the origin for times  less than 2 ns is 
  due primarily to the electrons movement while after the $2\, \textrm{ns}$ the distribution is due to the ion drifting towards to the mesh.  
  }\label{fig:graph_current1}
\end{figure}

\noindent Assuming that for a "classical Micromegas" $d=130\, \upmu \textrm{m},\ u_\textrm{n} = 10\, \textrm{cm}/\upmu \textrm{s},\ u_\textrm{p} = 5\times 10^2 \, \textrm{m/s}, \ \alpha = 550\, \textrm{cm}^{-1}\ q_0 = e = 1.60\times 10^{-19}\, \textrm{C},\ 
\beta = 0.995 \approx 1,00$. We calculated some quantities for the mesh. In addition we consider homogeneous auxiliary field. We give the results: The total
(negative) strip charge is  $Q_{\textrm{nt}} = 2,867\times 10^{-17}\, \textrm{C} = 179 e$, while the total (positive) ion charge is $Q_{\textrm{pt}} = 1,745\times 10^{-16}\, \textrm{C} = 1090 e$. The ratio of these charges is 
$Q_{\textrm{nt}}/Q_{\textrm{pt}} = 6,1$ and the charge amplification is 
$(Q_{\textrm{pt}}+Q_{\textrm{nt}})/Q_{\textrm{pt}} =1269$. Figure~\ref{fig:graph_current1} shows the total auxiliary current, $i_\textrm{t}$, versus 
time. 

%% file: figures/outerlattice.pdf_tex
\begingroup%
  \makeatletter%
  \providecommand\color[2][]{%
    \errmessage{(Inkscape) Color is used for the text in Inkscape, but the package 'color.sty' is not loaded}%
    \renewcommand\color[2][]{}%
  }%
  \providecommand\transparent[1]{%
    \errmessage{(Inkscape) Transparency is used (non-zero) for the text in Inkscape, but the package 'transparent.sty' is not loaded}%
    \renewcommand\transparent[1]{}%
  }%
  \providecommand\rotatebox[2]{#2}%
  \ifx\svgwidth\undefined%
    \setlength{\unitlength}{223.50630493bp}%
    \ifx\svgscale\undefined%
      \relax%
    \else%
      \setlength{\unitlength}{\unitlength * \real{\svgscale}}%
    \fi%
  \else%
    \setlength{\unitlength}{\svgwidth}%
  \fi%
  \global\let\svgwidth\undefined%
  \global\let\svgscale\undefined%
  \makeatother%
  \begin{picture}(1,1.01421505)%
    \put(0,0){\includegraphics[width=\unitlength]{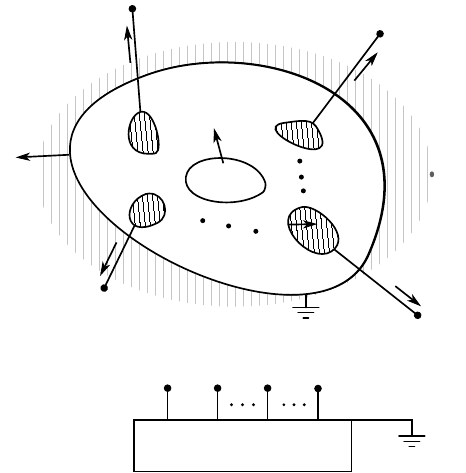}}%
    \put(0.45918937,0.74790115){\color[rgb]{0,0,0}\makebox(0,0)[lb]{\smash{$\bm{u}(\bm{x}, t)$}}}%
    \put(0.70448807,0.35044552){\color[rgb]{0,0,0}\makebox(0,0)[b]{\smash{$0$}}}%
    \put(0.25009156,0.36164475){\color[rgb]{0,0,0}\makebox(0,0)[b]{\smash{$N$}}}%
    \put(0.2751402,0.32931118){\color[rgb]{0,0,0}\makebox(0,0)[b]{\smash{$v_N(t)$}}}%
    \put(0.20518692,0.46883591){\color[rgb]{0,0,0}\makebox(0,0)[b]{\smash{$i_N$}}}%
    \put(0.06337384,0.63684836){\color[rgb]{0,0,0}\makebox(0,0)[b]{\smash{$\bm{n}$}}}%
    \put(0.23382146,0.89835402){\color[rgb]{0,0,0}\makebox(0,0)[b]{\smash{$i_1$}}}%
    \put(0.24066343,0.99106134){\color[rgb]{0,0,0}\makebox(0,0)[b]{\smash{1}}}%
    \put(0.34672652,0.98074694){\color[rgb]{0,0,0}\makebox(0,0)[b]{\smash{$v_{\textrm{1}}(t)$}}}%
    \put(0.81531441,0.85141864){\color[rgb]{0,0,0}\makebox(0,0)[b]{\smash{$i_2$}}}%
    \put(0.73460926,0.97674407){\color[rgb]{0,0,0}\makebox(0,0)[b]{\smash{2}}}%
    \put(0.87163219,0.95296089){\color[rgb]{0,0,0}\makebox(0,0)[b]{\smash{$v_{\textrm{2}}(t)$}}}%
    \put(0.88240728,0.29448198){\color[rgb]{0,0,0}\makebox(0,0)[b]{\smash{$v_{\ell}(t)$ $\ \ \ell$}}}%
    \put(0.89116387,0.39549786){\color[rgb]{0,0,0}\makebox(0,0)[b]{\smash{$i_{\ell}$}}}%
    \put(0.86823835,0.61388182){\color[rgb]{0,0,0}\makebox(0,0)[b]{\smash{$S_0$}}}%
    \put(0.73938292,0.71410271){\color[rgb]{0,0,0}\makebox(0,0)[b]{\smash{$S_2$}}}%
    \put(0.74654155,0.54229547){\color[rgb]{0,0,0}\makebox(0,0)[b]{\smash{$S_{\ell}$}}}%
    \put(0.36713391,0.5065023){\color[rgb]{0,0,0}\makebox(0,0)[b]{\smash{$S_N$}}}%
    \put(0.23111985,0.72126134){\color[rgb]{0,0,0}\makebox(0,0)[b]{\smash{$S_1$}}}%
    \put(0.55079917,0.42777796){\color[rgb]{0,0,0}\makebox(0,0)[b]{\smash{${\it \Omega}$}}}%
    \put(0.60027142,0.52946884){\color[rgb]{0,0,0}\makebox(0,0)[b]{\smash{$\bm{n}$}}}%
    \put(0.52607279,0.04666199){\color[rgb]{0,0,0}\makebox(0,0)[b]{\smash{External network}}}%
    \put(0.33440792,0.19496974){\color[rgb]{0,0,0}\makebox(0,0)[b]{\smash{1}}}%
    \put(0.44178744,0.19496974){\color[rgb]{0,0,0}\makebox(0,0)[b]{\smash{2}}}%
    \put(0.54916696,0.19496974){\color[rgb]{0,0,0}\makebox(0,0)[b]{\smash{$\ell$}}}%
    \put(0.65654649,0.19496974){\color[rgb]{0,0,0}\makebox(0,0)[b]{\smash{$N$}}}%
    \put(0.48304083,0.61465264){\color[rgb]{0,0,0}\makebox(0,0)[b]{\smash{$\rho(\bm{x}, t)$}}}%
  \end{picture}%
\endgroup%

%% file: figures/equivcirc.pdf_tex
\begingroup%
  \makeatletter%
  \providecommand\color[2][]{%
    \errmessage{(Inkscape) Color is used for the text in Inkscape, but the package 'color.sty' is not loaded}%
    \renewcommand\color[2][]{}%
  }%
  \providecommand\transparent[1]{%
    \errmessage{(Inkscape) Transparency is used (non-zero) for the text in Inkscape, but the package 'transparent.sty' is not loaded}%
    \renewcommand\transparent[1]{}%
  }%
  \providecommand\rotatebox[2]{#2}%
  \ifx\svgwidth\undefined%
    \setlength{\unitlength}{277.16784668bp}%
    \ifx\svgscale\undefined%
      \relax%
    \else%
      \setlength{\unitlength}{\unitlength * \real{\svgscale}}%
    \fi%
  \else%
    \setlength{\unitlength}{\svgwidth}%
  \fi%
  \global\let\svgwidth\undefined%
  \global\let\svgscale\undefined%
  \makeatother%
  \begin{picture}(1,0.59350918)%
    \put(0,0){\includegraphics[width=\unitlength]{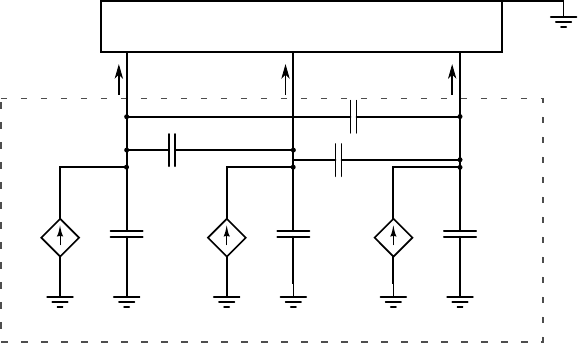}}%
    \put(0.51432942,0.539422){\color[rgb]{0,0,0}\makebox(0,0)[b]{\smash{External Network}}}%
    \put(0.51432942,0.00833584){\color[rgb]{0,0,0}\makebox(0,0)[b]{\smash{Internal Network}}}%
    \put(0.16394341,0.43506999){\color[rgb]{0,0,0}\makebox(0,0)[b]{\smash{$i_1$}}}%
    \put(0.26785155,0.43506999){\color[rgb]{0,0,0}\makebox(0,0)[b]{\smash{$v_1(t)$}}}%
    \put(0.45257718,0.43506999){\color[rgb]{0,0,0}\makebox(0,0)[b]{\smash{$i_2$}}}%
    \put(0.55648534,0.43506999){\color[rgb]{0,0,0}\makebox(0,0)[b]{\smash{$v_2(t)$}}}%
    \put(0.74121095,0.43506999){\color[rgb]{0,0,0}\makebox(0,0)[b]{\smash{$i_3$}}}%
    \put(0.84511911,0.43506999){\color[rgb]{0,0,0}\makebox(0,0)[b]{\smash{$v_3(t)$}}}%
    \put(0.1993019,0.27261459){\color[rgb]{0,0,0}\makebox(0,0)[b]{\smash{1}}}%
    \put(0.48793561,0.27261459){\color[rgb]{0,0,0}\makebox(0,0)[b]{\smash{2}}}%
    \put(0.78234206,0.27261459){\color[rgb]{0,0,0}\makebox(0,0)[b]{\smash{3}}}%
    \put(0.03183941,0.17194811){\color[rgb]{0,0,0}\makebox(0,0)[b]{\smash{$I_1$}}}%
    \put(0.32047317,0.17194811){\color[rgb]{0,0,0}\makebox(0,0)[b]{\smash{$I_2$}}}%
    \put(0.60910692,0.17194811){\color[rgb]{0,0,0}\makebox(0,0)[b]{\smash{$I_3$}}}%
    \put(0.18192897,0.15463008){\color[rgb]{0,0,0}\makebox(0,0)[b]{\smash{$C_{11}$}}}%
    \put(0.47056271,0.15463008){\color[rgb]{0,0,0}\makebox(0,0)[b]{\smash{$C_{22}$}}}%
    \put(0.75919648,0.15463008){\color[rgb]{0,0,0}\makebox(0,0)[b]{\smash{$C_{33}$}}}%
    \put(0.29738245,0.27585624){\color[rgb]{0,0,0}\makebox(0,0)[b]{\smash{$C_{12}$}}}%
    \put(0.661061,0.3566737){\color[rgb]{0,0,0}\makebox(0,0)[b]{\smash{$C_{13}$}}}%
    \put(0.58601622,0.25853819){\color[rgb]{0,0,0}\makebox(0,0)[b]{\smash{$C_{23}$}}}%
  \end{picture}%
\endgroup%

%% file: figures/det2cond.pdf_tex
\begingroup%
  \makeatletter%
  \providecommand\color[2][]{%
    \errmessage{(Inkscape) Color is used for the text in Inkscape, but the package 'color.sty' is not loaded}%
    \renewcommand\color[2][]{}%
  }%
  \providecommand\transparent[1]{%
    \errmessage{(Inkscape) Transparency is used (non-zero) for the text in Inkscape, but the package 'transparent.sty' is not loaded}%
    \renewcommand\transparent[1]{}%
  }%
  \providecommand\rotatebox[2]{#2}%
  \ifx\svgwidth\undefined%
    \setlength{\unitlength}{226.65788607bp}%
    \ifx\svgscale\undefined%
      \relax%
    \else%
      \setlength{\unitlength}{\unitlength * \real{\svgscale}}%
    \fi%
  \else%
    \setlength{\unitlength}{\svgwidth}%
  \fi%
  \global\let\svgwidth\undefined%
  \global\let\svgscale\undefined%
  \makeatother%
  \begin{picture}(1,0.78016779)%
    \put(0,0){\includegraphics[width=\unitlength]{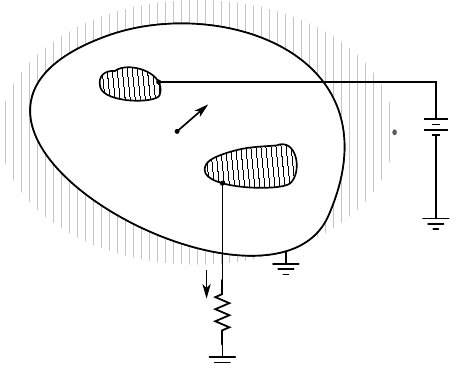}}%
    \put(0.53122662,0.17920895){\color[rgb]{0,0,0}\makebox(0,0)[b]{\smash{$+$}}}%
    \put(0.53122662,0.03587361){\color[rgb]{0,0,0}\makebox(0,0)[b]{\smash{$-$}}}%
    \put(0.53122662,0.10878946){\color[rgb]{0,0,0}\makebox(0,0)[b]{\smash{$R$}}}%
    \put(0.59856545,0.14708697){\color[rgb]{0,0,0}\makebox(0,0)[b]{\smash{$v_2(t)$}}}%
    \put(0.36390105,0.17763446){\color[rgb]{0,0,0}\makebox(0,0)[b]{\smash{$i$}}}%
    \put(0.34929141,0.47206559){\color[rgb]{0,0,0}\makebox(0,0)[b]{\smash{$q$}}}%
    \put(0.46176588,0.51722273){\color[rgb]{0,0,0}\makebox(0,0)[b]{\smash{$\bm{u}$}}}%
    \put(0.32341756,0.64353778){\color[rgb]{0,0,0}\makebox(0,0)[b]{\smash{1}}}%
    \put(0.4673506,0.61949033){\color[rgb]{0,0,0}\makebox(0,0)[b]{\smash{$v_1(t)=V_a$}}}%
    \put(0.63336621,0.46931098){\color[rgb]{0,0,0}\makebox(0,0)[b]{\smash{2}}}%
    \put(0.87071052,0.46185873){\color[rgb]{0,0,0}\makebox(0,0)[b]{\smash{$V_a$}}}%
    \put(0.96889068,0.54628197){\color[rgb]{0,0,0}\makebox(0,0)[b]{\smash{$+$}}}%
    \put(0.96889068,0.44411933){\color[rgb]{0,0,0}\makebox(0,0)[b]{\smash{$-$}}}%
  \end{picture}%
\endgroup%

%% file: figures/equivdet2cond.pdf_tex
\begingroup%
  \makeatletter%
  \providecommand\color[2][]{%
    \errmessage{(Inkscape) Color is used for the text in Inkscape, but the package 'color.sty' is not loaded}%
    \renewcommand\color[2][]{}%
  }%
  \providecommand\transparent[1]{%
    \errmessage{(Inkscape) Transparency is used (non-zero) for the text in Inkscape, but the package 'transparent.sty' is not loaded}%
    \renewcommand\transparent[1]{}%
  }%
  \providecommand\rotatebox[2]{#2}%
  \ifx\svgwidth\undefined%
    \setlength{\unitlength}{245.84406738bp}%
    \ifx\svgscale\undefined%
      \relax%
    \else%
      \setlength{\unitlength}{\unitlength * \real{\svgscale}}%
    \fi%
  \else%
    \setlength{\unitlength}{\svgwidth}%
  \fi%
  \global\let\svgwidth\undefined%
  \global\let\svgscale\undefined%
  \makeatother%
  \begin{picture}(1,0.51965705)%
    \put(0,0){\includegraphics[width=\unitlength]{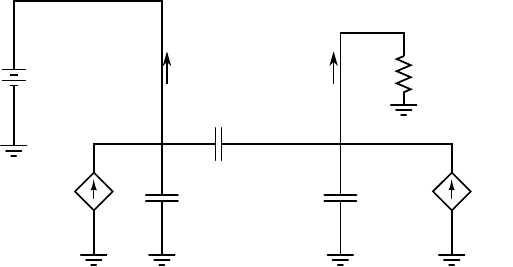}}%
    \put(0.08611563,0.36302977){\color[rgb]{0,0,0}\makebox(0,0)[b]{\smash{$V_a$}}}%
    \put(0.09913201,0.13687016){\color[rgb]{0,0,0}\makebox(0,0)[b]{\smash{$I_1$}}}%
    \put(0.96472133,0.13687016){\color[rgb]{0,0,0}\makebox(0,0)[b]{\smash{$I_2$}}}%
    \put(0.72601833,0.09254259){\color[rgb]{0,0,0}\makebox(0,0)[b]{\smash{$C_{22}$}}}%
    \put(0.37754643,0.09254259){\color[rgb]{0,0,0}\makebox(0,0)[b]{\smash{$C_{11}$}}}%
    \put(0.46359067,0.16941111){\color[rgb]{0,0,0}\makebox(0,0)[b]{\smash{$C_{12}$}}}%
    \put(0.83455752,0.36465682){\color[rgb]{0,0,0}\makebox(0,0)[b]{\smash{$R$}}}%
    \put(0.61327905,0.3776732){\color[rgb]{0,0,0}\makebox(0,0)[b]{\smash{$i_2$}}}%
    \put(0.372476,0.3776732){\color[rgb]{0,0,0}\makebox(0,0)[b]{\smash{$i_1$}}}%
  \end{picture}%
\endgroup%

%% file: figures/crosssectcond.pdf_tex
\begingroup%
  \makeatletter%
  \providecommand\color[2][]{%
    \errmessage{(Inkscape) Color is used for the text in Inkscape, but the package 'color.sty' is not loaded}%
    \renewcommand\color[2][]{}%
  }%
  \providecommand\transparent[1]{%
    \errmessage{(Inkscape) Transparency is used (non-zero) for the text in Inkscape, but the package 'transparent.sty' is not loaded}%
    \renewcommand\transparent[1]{}%
  }%
  \providecommand\rotatebox[2]{#2}%
  \ifx\svgwidth\undefined%
    \setlength{\unitlength}{224.9958287bp}%
    \ifx\svgscale\undefined%
      \relax%
    \else%
      \setlength{\unitlength}{\unitlength * \real{\svgscale}}%
    \fi%
  \else%
    \setlength{\unitlength}{\svgwidth}%
  \fi%
  \global\let\svgwidth\undefined%
  \global\let\svgscale\undefined%
  \makeatother%
  \begin{picture}(1,0.42221777)%
    \put(0,0){\includegraphics[width=\unitlength]{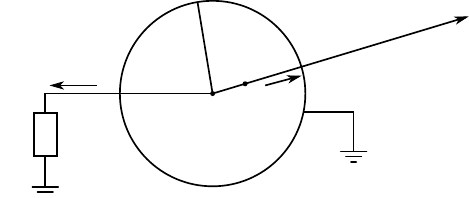}}%
    \put(0.7998764,0.18223534){\color[rgb]{0,0,0}\makebox(0,0)[b]{\smash{0}}}%
    \put(0.18202247,0.25867901){\color[rgb]{0,0,0}\makebox(0,0)[b]{\smash{$i(t)$}}}%
    \put(0.03979761,0.18756655){\color[rgb]{0,0,0}\makebox(0,0)[b]{\smash{$v(t)$}}}%
    \put(0.46785305,0.37052407){\color[rgb]{0,0,0}\makebox(0,0)[b]{\smash{$b$}}}%
    \put(0.60785305,0.37052407){\color[rgb]{0,0,0}\makebox(0,0)[b]{\smash{$2$}}}%
    \put(0.42079114,0.26029983){\color[rgb]{0,0,0}\makebox(0,0)[b]{\smash{1}}}%
    \put(0.40766883,0.18246012){\color[rgb]{0,0,0}\makebox(0,0)[b]{\smash{+}}}%
    \put(0.46663992,0.18694496){\color[rgb]{0,0,0}\makebox(0,0)[b]{\smash{$a$}}}%
    \put(0.48569165,0.25747019){\color[rgb]{0,0,0}\makebox(0,0)[b]{\smash{$r$}}}%
    \put(0.54207665,0.26918888){\color[rgb]{0,0,0}\makebox(0,0)[b]{\smash{$q$}}}%
    \put(0.6100657,0.20215859){\color[rgb]{0,0,0}\makebox(0,0)[b]{\smash{$u(r)$}}}%
  \end{picture}%
\endgroup%

%% file: figures/equivcrosssectcond.pdf_tex
\begingroup%
  \makeatletter%
  \providecommand\color[2][]{%
    \errmessage{(Inkscape) Color is used for the text in Inkscape, but the package 'color.sty' is not loaded}%
    \renewcommand\color[2][]{}%
  }%
  \providecommand\transparent[1]{%
    \errmessage{(Inkscape) Transparency is used (non-zero) for the text in Inkscape, but the package 'transparent.sty' is not loaded}%
    \renewcommand\transparent[1]{}%
  }%
  \providecommand\rotatebox[2]{#2}%
  \ifx\svgwidth\undefined%
    \setlength{\unitlength}{198.29251709bp}%
    \ifx\svgscale\undefined%
      \relax%
    \else%
      \setlength{\unitlength}{\unitlength * \real{\svgscale}}%
    \fi%
  \else%
    \setlength{\unitlength}{\svgwidth}%
  \fi%
  \global\let\svgwidth\undefined%
  \global\let\svgscale\undefined%
  \makeatother%
  \begin{picture}(1,0.49396094)%
    \put(0,0){\includegraphics[width=\unitlength]{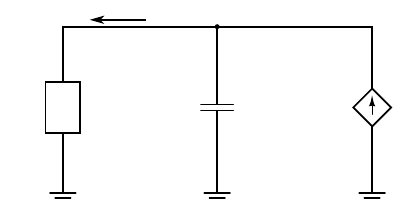}}%
    \put(0.96840408,0.01371836){\color[rgb]{0,0,0}\makebox(0,0)[b]{\smash{0}}}%
    \put(0.97647297,0.2235094){\color[rgb]{0,0,0}\makebox(0,0)[b]{\smash{$I$}}}%
    \put(0.62144192,0.2235094){\color[rgb]{0,0,0}\makebox(0,0)[b]{\smash{$C_d$}}}%
    \put(0.29868642,0.46557602){\color[rgb]{0,0,0}\makebox(0,0)[b]{\smash{$I(t)$}}}%
    \put(0.04515701,0.22221869){\color[rgb]{0,0,0}\makebox(0,0)[b]{\smash{$v(t)$}}}%
    \put(0.52312957,0.45087188){\color[rgb]{0,0,0}\makebox(0,0)[b]{\smash{1}}}%
  \end{picture}%
\endgroup%

%% file: figures/indicativesignal.pdf_tex
\begingroup%
  \makeatletter%
  \providecommand\color[2][]{%
    \errmessage{(Inkscape) Color is used for the text in Inkscape, but the package 'color.sty' is not loaded}%
    \renewcommand\color[2][]{}%
  }%
  \providecommand\transparent[1]{%
    \errmessage{(Inkscape) Transparency is used (non-zero) for the text in Inkscape, but the package 'transparent.sty' is not loaded}%
    \renewcommand\transparent[1]{}%
  }%
  \providecommand\rotatebox[2]{#2}%
  \ifx\svgwidth\undefined%
    \setlength{\unitlength}{292.55599442bp}%
    \ifx\svgscale\undefined%
      \relax%
    \else%
      \setlength{\unitlength}{\unitlength * \real{\svgscale}}%
    \fi%
  \else%
    \setlength{\unitlength}{\svgwidth}%
  \fi%
  \global\let\svgwidth\undefined%
  \global\let\svgscale\undefined%
  \makeatother%
  \begin{picture}(1,0.51748565)%
    \put(0,0){\includegraphics[width=\unitlength]{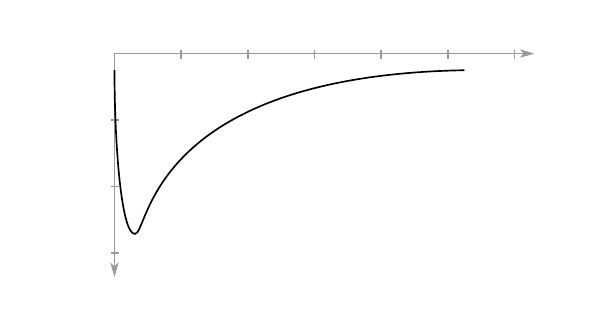}}%
    \put(0.16966126,0.41927236){\color[rgb]{0,0,0}\makebox(0,0)[rb]{\smash{$0$}}}%
    \put(0.15997251,0.0051793){\color[rgb]{0,0,0}\makebox(0,0)[b]{\smash{$\times 10^{-4}$}}}%
    \put(0.91257683,0.39775642){\color[rgb]{0,0,0}\makebox(0,0)[b]{\smash{$\times 10^{-6}$}}}%
    \put(0.16966126,0.30989158){\color[rgb]{0,0,0}\makebox(0,0)[rb]{\smash{$-1$}}}%
    \put(0.16966126,0.20051083){\color[rgb]{0,0,0}\makebox(0,0)[rb]{\smash{$-2$}}}%
    \put(0.16966126,0.09113006){\color[rgb]{0,0,0}\makebox(0,0)[rb]{\smash{$-3$}}}%
    \put(0.29736387,0.44606788){\color[rgb]{0,0,0}\makebox(0,0)[b]{\smash{$0.2$}}}%
    \put(0.40674464,0.44606788){\color[rgb]{0,0,0}\makebox(0,0)[b]{\smash{$0.4$}}}%
    \put(0.51612537,0.44606788){\color[rgb]{0,0,0}\makebox(0,0)[b]{\smash{$0.6$}}}%
    \put(0.62550618,0.44606788){\color[rgb]{0,0,0}\makebox(0,0)[b]{\smash{$0.8$}}}%
    \put(0.73488695,0.44606788){\color[rgb]{0,0,0}\makebox(0,0)[b]{\smash{$1$}}}%
    \put(0.84426776,0.44606788){\color[rgb]{0,0,0}\makebox(0,0)[b]{\smash{$1.2$}}}%
    \put(0.07968226,0.24736786){\color[rgb]{0,0,0}\makebox(0,0)[b]{\smash{$v(t)/\mathrm{V}$}}}%
    \put(0.62531482,0.49824655){\color[rgb]{0,0,0}\makebox(0,0)[b]{\smash{$t/$s}}}%
  \end{picture}%
\endgroup%

%% file: figures/micromegas_geo.pdf_tex
\begingroup%
  \makeatletter%
  \providecommand\color[2][]{%
    \errmessage{(Inkscape) Color is used for the text in Inkscape, but the package 'color.sty' is not loaded}%
    \renewcommand\color[2][]{}%
  }%
  \providecommand\transparent[1]{%
    \errmessage{(Inkscape) Transparency is used (non-zero) for the text in Inkscape, but the package 'transparent.sty' is not loaded}%
    \renewcommand\transparent[1]{}%
  }%
  \providecommand\rotatebox[2]{#2}%
  \ifx\svgwidth\undefined%
    \setlength{\unitlength}{279.64109802bp}%
    \ifx\svgscale\undefined%
      \relax%
    \else%
      \setlength{\unitlength}{\unitlength * \real{\svgscale}}%
    \fi%
  \else%
    \setlength{\unitlength}{\svgwidth}%
  \fi%
  \global\let\svgwidth\undefined%
  \global\let\svgscale\undefined%
  \makeatother%
  \begin{picture}(1,0.3171692)%
    \put(0,0){\includegraphics[width=\unitlength,page=1]{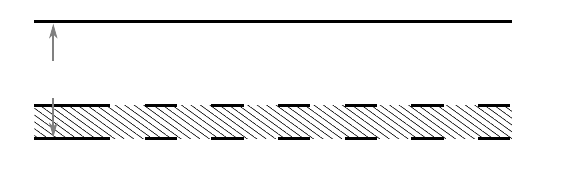}}%
    \put(0.09200706,0.16896122){\color[rgb]{0,0,0}\makebox(0,0)[b]{\smash{$d$}}}%
    \put(0.28148525,0.10036127){\color[rgb]{0,0,0}\makebox(0,0)[b]{\smash{$\varepsilon_{\textrm{mr}}\quad$ insulator}}}%
    \put(0.27527887,0.15117012){\color[rgb]{0,0,0}\makebox(0,0)[b]{\smash{$\varepsilon_{\textrm{r}}$}}}%
    \put(0.44324716,0.1861856){\color[rgb]{0,0,0}\makebox(0,0)[b]{\smash{$E_\textrm{g}$}}}%
    \put(0.58056604,0.19762883){\color[rgb]{0,0,0}\makebox(0,0)[b]{\smash{$q$}}}%
    \put(0.58056604,0.23768016){\color[rgb]{0,0,0}\makebox(0,0)[b]{\smash{A}}}%
    \put(0,0){\includegraphics[width=\unitlength,page=2]{micromegas_geo.pdf}}%
    \put(0.63445155,0.10703404){\color[rgb]{0,0,0}\makebox(0,0)[b]{\smash{resistive strips}}}%
    \put(0.63445155,0.04336218){\color[rgb]{0,0,0}\makebox(0,0)[b]{\smash{readout strips}}}%
    \put(0,0){\includegraphics[width=\unitlength,page=3]{micromegas_geo.pdf}}%
    \put(0.83547189,0.20300329){\color[rgb]{0,0,0}\makebox(0,0)[b]{\smash{$d_\textrm{g}$}}}%
    \put(0.30377103,0.22953995){\color[rgb]{0,0,0}\makebox(0,0)[b]{\smash{amplification region}}}%
    \put(1.00214977,0.10131316){\color[rgb]{0,0,0}\makebox(0,0)[rb]{\smash{$d-d_\textrm{g}$}}}%
    \put(0.81882004,0.28930271){\color[rgb]{0,0,0}\makebox(0,0)[b]{\smash{mesh}}}%
    \put(0,0){\includegraphics[width=\unitlength,page=4]{micromegas_geo.pdf}}%
    \put(0.92833646,0.24381224){\color[rgb]{0,0,0}\makebox(0,0)[b]{\smash{$x$}}}%
    \put(0.14158935,0.00378417){\color[rgb]{0,0,0}\makebox(0,0)[lb]{\smash{$z$}}}%
    \put(0,0){\includegraphics[width=\unitlength,page=5]{micromegas_geo.pdf}}%
    \put(0.13507163,0.2933022){\color[rgb]{0,0,0}\makebox(0,0)[b]{\smash{O}}}%
  \end{picture}%
\endgroup%

%% file: figures/chdistr_shower.pdf_tex
\begingroup%
  \makeatletter%
  \providecommand\color[2][]{%
    \errmessage{(Inkscape) Color is used for the text in Inkscape, but the package 'color.sty' is not loaded}%
    \renewcommand\color[2][]{}%
  }%
  \providecommand\transparent[1]{%
    \errmessage{(Inkscape) Transparency is used (non-zero) for the text in Inkscape, but the package 'transparent.sty' is not loaded}%
    \renewcommand\transparent[1]{}%
  }%
  \providecommand\rotatebox[2]{#2}%
  \ifx\svgwidth\undefined%
    \setlength{\unitlength}{211.72581482bp}%
    \ifx\svgscale\undefined%
      \relax%
    \else%
      \setlength{\unitlength}{\unitlength * \real{\svgscale}}%
    \fi%
  \else%
    \setlength{\unitlength}{\svgwidth}%
  \fi%
  \global\let\svgwidth\undefined%
  \global\let\svgscale\undefined%
  \makeatother%
  \begin{picture}(1,0.55469863)%
    \put(0,0){\includegraphics[width=\unitlength,page=1]{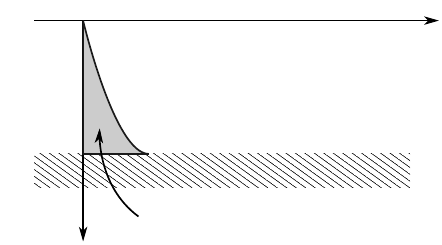}}%
    \put(0.45096483,0.03853192){\color[rgb]{0,0,0}\makebox(0,0)[b]{\smash{$\fr{\mathrm{d}q_{\textrm{p}}}{\mathrm{d}z}= f(z)$}}}%
    \put(0.22849528,0.00499799){\color[rgb]{0,0,0}\makebox(0,0)[b]{\smash{$z$}}}%
    \put(0,0){\includegraphics[width=\unitlength,page=2]{chdistr_shower.pdf}}%
    \put(0.92153216,0.524376){\color[rgb]{0,0,0}\makebox(0,0)[b]{\smash{$-$}}}%
    \put(0.92153216,0.2214402){\color[rgb]{0,0,0}\makebox(0,0)[b]{\smash{$+$}}}%
    \put(0.69264104,0.34534336){\color[rgb]{0,0,0}\makebox(0,0)[b]{\smash{$d_\textrm{g}$}}}%
    \put(0.37682611,0.37508497){\color[rgb]{0,0,0}\makebox(0,0)[b]{\smash{$q_{\textrm{p}}$}}}%
    \put(0.38909634,0.25864327){\color[rgb]{0,0,0}\makebox(0,0)[b]{\smash{$u_{\textrm{n}}$}}}%
    \put(0.52040797,0.44309747){\color[rgb]{0,0,0}\makebox(0,0)[b]{\smash{$u_{\textrm{p}}$}}}%
    \put(0.52796492,0.32974332){\color[rgb]{0,0,0}\makebox(0,0)[b]{\smash{$q_{\textrm{n}}$}}}%
    \put(0.80762181,0.16268057){\color[rgb]{0,0,0}\makebox(0,0)[b]{\smash{resistive strips}}}%
    \put(0.80817807,0.51729136){\color[rgb]{0,0,0}\makebox(0,0)[b]{\smash{mesh}}}%
    \put(0.97112466,0.46061429){\color[rgb]{0,0,0}\makebox(0,0)[b]{\smash{$x$}}}%
    \put(0.18899677,0.52317581){\color[rgb]{0,0,0}\makebox(0,0)[b]{\smash{O}}}%
    \put(0.02889203,0.3094754){\color[rgb]{0,0,0}\makebox(0,0)[b]{\smash{$y$}}}%
    \put(0,0){\includegraphics[width=\unitlength,page=3]{chdistr_shower.pdf}}%
    \put(0.12587025,0.30852687){\color[rgb]{0,0,0}\makebox(0,0)[b]{\smash{$d$}}}%
    \put(0,0){\includegraphics[width=\unitlength,page=4]{chdistr_shower.pdf}}%
    \put(0.80762181,0.08786993){\color[rgb]{0,0,0}\makebox(0,0)[b]{\smash{readout strips}}}%
  \end{picture}%
\endgroup%

%% file: figures/graph10.pdf_tex
\begingroup%
  \makeatletter%
  \providecommand\color[2][]{%
    \errmessage{(Inkscape) Color is used for the text in Inkscape, but the package 'color.sty' is not loaded}%
    \renewcommand\color[2][]{}%
  }%
  \providecommand\transparent[1]{%
    \errmessage{(Inkscape) Transparency is used (non-zero) for the text in Inkscape, but the package 'transparent.sty' is not loaded}%
    \renewcommand\transparent[1]{}%
  }%
  \providecommand\rotatebox[2]{#2}%
  \ifx\svgwidth\undefined%
    \setlength{\unitlength}{189.05860871bp}%
    \ifx\svgscale\undefined%
      \relax%
    \else%
      \setlength{\unitlength}{\unitlength * \real{\svgscale}}%
    \fi%
  \else%
    \setlength{\unitlength}{\svgwidth}%
  \fi%
  \global\let\svgwidth\undefined%
  \global\let\svgscale\undefined%
  \makeatother%
  \begin{picture}(1,0.56091872)%
    \put(0.012782,0.51588474){\color[rgb]{0,0,0}\makebox(0,0)[rb]{\smash{\(\drac{q_\textrm{p}}{z}\)}}}%
    \put(0,0){\includegraphics[width=\unitlength,page=1]{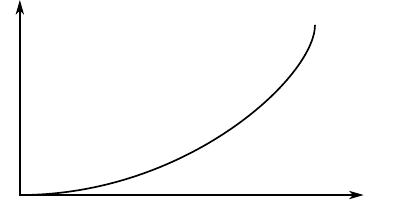}}%
    \put(0.91326977,0.00789334){\color[rgb]{0,0,0}\makebox(0,0)[lb]{\smash{\(z\)}}}%
    \put(0.80175848,0.00789334){\color[rgb]{0,0,0}\makebox(0,0)[b]{\smash{\(d_\textrm{g}\)}}}%
    \put(0.61566356,0.00789334){\color[rgb]{0,0,0}\makebox(0,0)[b]{\smash{\(z_1\)}}}%
    \put(0.44558928,0.00789334){\color[rgb]{0,0,0}\makebox(0,0)[b]{\smash{\(z_2\)}}}%
    \put(0.22343595,0.00789334){\color[rgb]{0,0,0}\makebox(0,0)[b]{\smash{\(z\)}}}%
    \put(0.04779606,0.00789334){\color[rgb]{0,0,0}\makebox(0,0)[b]{\smash{\(0\)}}}%
    \put(0,0){\includegraphics[width=\unitlength,page=2]{graph10.pdf}}%
    \put(0,0){\includegraphics[width=\unitlength,page=3]{graph10.pdf}}%
  \end{picture}%
\endgroup%

%% file: Acknowledgments.tex
\section*{Acknowledgments}

We would like to thank our former graduate students G. Iakovidis, S. Leontsinis, and K. Ntekas for carefully reading the 
manuscript and making valuable comments and also 
 Dan Ciubotaru \& Michele Renda from IFIN-HH Romania for their helpful 
 discussions 
 on the Micromegas analysis.\\

\noindent The present work was co-funded by the European Union (European Social Fund ESF) and Greek national funds through the Operational Program "Education and Lifelong Learning" of the National Strategic Reference Framework (NSRF) 2007-1013. ARISTEIA-1893-ATLAS MICROMEGAS.  